# Superconductivity and a van Hove singularity confined to the surface of a topological semimetal

**Authors:** Md Shafayat Hossain[1]*†, Rajibul Islam[2]*, Zi-Jia Cheng[1]*, Zahir Muhammad[3,4]*, Qi Zhang[1], Zurab Guguchia[5], Jonas A. Krieger[5], Brian Casas[6], Yu-Xiao Jiang[1], Maksim Litskevich[1], Xian P. Yang[1], Byunghoon Kim[1], Tyler A. Cochran[1], Ilias E. Perakis[2], Fei Xue[2], Mehdi Kargarian[7], Weisheng Zhao[3], Luis Balicas[5], M. Zahid Hasan[1,8]†

**Affiliations:**

[1]Laboratory for Topological Quantum Matter and Advanced Spectroscopy (B7), Department of Physics, Princeton University, Princeton, New Jersey, USA.

[2]Department of Physics, University of Alabama at Birmingham, Birmingham, Alabama 35294, USA.

[3]Hefei Innovation Research Institute, School of Microelectronics, Beihang University, Hefei 230013, P.R. China.

[4]National Key Laboratory of Spintronics, Hangzhou International Innovation Institute, Beihang University, Hangzhou 311115, P.R. China.

[5]PSI Center for Neutron and Muon Sciences CNM, 5232 Villigen PSI, Switzerland.

[6]National High Magnetic Field Laboratory, Tallahassee, Florida 32310, USA.

[7]Department of Physics, Sharif University of Technology, Tehran 14588-89694, Iran.

[8]Quantum Science Center at ORNL, Oak Ridge, TN, USA.

†Corresponding authors, E-mail: mdsh@princeton.edu; mzhasan@princeton.edu.

*These authors contributed equally to this work.

**Abstract:**

**The interplay between electronic topology and superconductivity is the subject of great current interest in condensed matter physics. For example, superconductivity induced on the surface of topological insulators is predicted to be triplet in nature, while the interplay between electronic correlations and topology may lead to unconventional superconductivity as in twisted bilayer graphene. Here, we unveil an unconventional two-dimensional superconducting state in the recently discovered Dirac nodal line semimetal $ZrAs_2$ which is exclusively confined to the top and bottom surfaces within the crystal's *ab* plane. As a remarkable consequence of this emergent state, we observe a Berezinskii–Kosterlitz–Thouless (BKT) transition, the hallmark of two-dimensional superconductivity. Notably, this is the first observation of a BKT transition on the surface of a three-dimensional system. Furthermore, employing angle-resolved photoemission spectroscopy and first-principles calculations, we find that these same surfaces also host a two-dimensional van Hove singularity near the Fermi energy. The proximity of van Hove singularity to the Fermi level leads to enhanced electronic correlations contributing to the stabilization of superconductivity at the surface of $ZrAs_2$, a unique phenomenon among topological semimetals. The surface-confined nature of the van Hove singularity, and associated superconductivity, realized for the first time, opens new avenues to explore the interplay between low-dimensional quantum topology, correlations, and superconductivity in a bulk material without resorting to the superconducting proximity effect.**

**Main Text:**

Condensed matter fermions can acquire topological characteristics through their energy-momentum dispersions, characterized, for example, by Dirac/Weyl-like band crossings[1-5]. With the advent of topological insulators[1,2], a diverse array of topological phases came to light, spanning from confined zero-dimensional Dirac crossings in momentum space, as in Dirac and Weyl semimetals[4-5], to extended one-dimensional Dirac/Weyl lines or even loops, detected in nodal line semimetals[6-11]. These topological phases possess low-energy excitations or quasiparticles whose behavior mirrors those of relativistic particles, offering avenues for studying elusive phenomena in high-energy physics while materializing novel quantum effects not yet observed in known compounds[1-4]. In addition, linearly dispersing bands and related quasiparticles can be harvested to stabilize unique ground states, as exemplified by the topological insulators whose linearly dispersing surface bands were predicted to lead to interfacial triplet superconductivity when proximitized with conventional superconductors[12]. In general, and as exemplified by the transistor, interfacial states are of great relevance both scientifically and technologically. Nevertheless, relatively little has been explored concerning interfaces involving the surface states in correspondence with bulk topological electronic states.

Recently, considerable attention has been focused on topological materials exhibiting electronic correlations and electronic order. Plausible ways to trigger correlations on surface states are via flatband-type drumhead states, as predicted for nodal line semimetals where a continuous line of Dirac nodes is observed[11,13-22], or the presence of van Hove singularities (vHss) near the Fermi level, as observed in Kagome superconductors[23-26]. In both scenarios, the drumhead states or vHss lead to pronounced peaks in the density of states, making these systems susceptible to the development of electronic order such as superconductivity, magnetism, or charge order[17-20, 24-26]. Superconducting nodal line compounds, where superconductivity condenses from torus-shaped Fermi surfaces derived from the nodal lines, are predicted to harbor topological crystalline and second-order topological superconductivity[20]. Although several nodal line materials were recently discovered[27-41], with some even exhibiting superconductivity[39-41], no direct evidence for the unconventional nature of these superconducting states has been reported. The same applies to Kagome superconductors, where the intertwining between vHss and charge order with superconductivity has been suggested[26], but no clear consensus on the nature of the pairing mechanism has emerged[42-45]. Here, we present the discovery of two-dimensional superconductivity confined exclusively to the *ab*-plane of the nodal line semimetal, ZrAs$_2$. Notably, our momentum space spectroscopic studies of ZrAs$_2$ reveal a two-dimensional vHs near the Fermi energy, confined to the same surfaces. Such proximity is likely to enhance electronic correlations contributing to the stabilization of superconductivity in the surface of ZrAs$_2$. Thus, ZrAs$_2$ potentially offers a natural solution to the stabilization of unconventional two-dimensional superconductivity in topological compounds, namely the coexistence between Dirac like quasiparticles and vHs at the surface state accompanied by a superconducting state.

We begin our discussion with an examination of the electronic band structure of ZrAs$_2$. ZrAs$_2$ belongs to the nonsymmorphic space group *Pnma* (No. 62, $D_{2h}^{16}$) (Fig. 1**a**) which is known to host Dirac nodal lines[22,46]; the sample characterization data can be found in Extended Figs. 1- 3. Figure 1**b** illustrates the bulk and (001) surface projected Brillouin zone. In Fig. 1**c**, we present the electronic band structure (in the absence of spin-orbit coupling), revealing metallic behavior with both electron- and hole-like bands crossing the Fermi level. The partially filled bulk bands yields two hole-like Fermi surface pockets ($FS_1^h$ and $FS_2^h$) and one electron-like Fermi surface pocket ($FS_1^e$) (Figs. 1**d** and 1**e**). This is consistent with our experimental findings exposing three peaks in the FFT spectrum from quantum oscillatory measurements (see Extended Figs. 4, 5). Examining the energy bands along the X → Γ, Γ → Z, and X → U → Z directions, we identify multiple band crossings, labeled as NL1, NL2, NL3, and NL4 between the magenta and cyan bands and NL5 between the green and magenta bands as shown in Fig. 1**c**. The electronic band structure calculations characterize ZrAs$_2$ as a nodal line semimetal with four nodal loops protected by the mirror $M_y$ and glide mirror ($G_x$ and $G_z$) planes[22]. Figure 1**f** plots the nodal loops/lines in the $k_x$-$k_y$-$k_z$ space, with black dots marking the nodal points along high-symmetry directions; the color bar shows the location of the nodes in energy. The distribution of the nodes (red lines) in the $k_z = 0$ plane is illustrated in Fig. 1**g**. A pair of concentric intersecting coplanar ellipses from half-filled bands form a butterfly-like nodal loop around the U $(\pi, 0, \pi)$ -point (Fig. 1**h**), as previously suggested[22].

To experimentally confirm the nodal features within the electronic band structure, we conducted angle-resolved photoemission spectroscopy (ARPES) measurements, which directly visualize the electronic bands. In Extended Fig. 6, we present ARPES data alongside an *ab-initio* energy-momentum cut along the Y → Γ→Y path. ARPES energy-momentum (*E-k*) cut along the $\overline{Y} - \overline{\Gamma} - \overline{Y}$ path (Extended Fig. 6**b**) reveals two sets of band crossings: one located at $E - E_f = -0.29$ eV and another one near the Fermi level. These crossings align closely with the *ab-initio* calculations (Extended Fig. 6**c**), despite some bands being suppressed by matrix element effects or $k_z$ broadening. These two crossings correspond to nodal lines NL5 and NL1, respectively. The excellent agreement between ARPES and the DFT calculations provides evidence for the existence of nodal lines in ZrAs$_2$. It is worth mentioning that, in the presence of spin-orbit coupling, the band crossings are lifted, leading to the emergence of a gap ranging from 2 to 58 meV (depending on the specific nodal feature), as depicted in Fig. 1**j**. The small gaps (*e.g.*, 10 meV and 2 meV for NL1 and NL5, respectively) fall below the ARPES energy resolution (approximately 15 meV) and, therefore, cannot be observed under the presence of the $k_z$ or extrinsic broadening and thus can be generally ignored[47,48]; see Extended Fig. 6**c**. Therefore, NL1 and NL5 can be effectively treated as a nodal line state under these conditions.

Having uncovered the band topology of ZrAs$_2$, we performed systematic electrical transport measurements. For this investigation, we selected a small, uniform, needle-shaped crystal and attached four electrical contacts on the *ab*-plane [(001) surface] of ZrAs$_2$ using silver epoxy; see Supplementary Information Section I and Supplementary Fig. 1 on how we determined the crystal plane prior to conducting the measurements. This configuration allows measurements of the longitudinal resistance ($R$) using the standard four-probe method. Figure 2**a** illustrates the temperature-dependence of $R$. At elevated temperatures, $R$ displays metallic behavior ($dR/dT > 0$), indicative of transport being dominated by phonon-scattering[49]. The residual resistivity ratio of approximately 35 indicates high sample quality and a significant variation in resistivity with temperature. Remarkably, the resistance undergoes a sudden drop to zero, marking the onset of superconductivity with a $T_c$ of 1.8 K, which is the highlight of our investigation. Notice that Zr becomes a type-I superconductor below $T_c$ of 0.6 K, while As only becomes superconducting under very high hydrostatic pressures[50,51]. Therefore, the observed superconducting state ought to be intrinsic to ZrAs$_2$. To investigate its characteristics, we probed the magnetoresistance well below $T_c$, at $T = 0.3$ K, while varying the magnetic field orientation, spanning from the *c*-axis [001] ($\theta = 0^o$) to the [100] direction ($\theta = 90^o$) of ZrAs$_2$ (Fig. 2**b**). A crucial observation obtained from the magnetoresistance traces in Fig. 2**b** is the upper critical magnetic field ($H_{c2}$), corresponding to the field at which the resistance attains 50% of its normal state value. In Fig. 2**c**, we chart $H_{c2}$ as a function of $\theta$. As the magnetic field direction progresses from the out-of-plane ($\theta = 0^o$) to in-plane ($\theta = 90^o$) orientation with respect to the (001)-plane of ZrAs$_2$, the transition to the (field-induced) normal state systematically shifts to higher fields, with the maximum $\mu_0 H_{c2}(\theta)$ recorded at $\theta = 90^o$ (Fig. 3**c**). The in-plane $\mu_0 H_{c2}$ is determined to be approximately 3.66 T, which is 11% larger than the weak coupling Pauli (or Clogston–Chandrasekhar) paramagnetic limit, $\mu_0 H_p = 1.84 \times T_c$ T/K= 3.3 T for BCS superconductors[52,53]. This observation underscores a superconducting anisotropy, captured by the Tinkham formula governing the angular dependence of $H_{c2}$ for a two-dimensional superconductor[54]: $\left|\frac{H_{c2}(\theta)\cos\theta}{H_{c2,\perp}}\right| + \left(\frac{H_{c2}(\theta)\sin\theta}{H_{c2,\|}}\right)^2 = 1$. Here, $\mu_0 H_{c2,\perp}$ and $\mu_0 H_{c2,\|}$ represent the upper critical magnetic fields for fields perpendicular and parallel to the sample plane, respectively. As depicted in Fig. 2**c**, our data aligns well with the Tinkham formula while revealing deviations (for near $\theta = 90°$) from the Ginzburg-Landau anisotropic mass model[52], which characterizes the angular dependence of $\mu_0 H_{c2}$ for a three-dimensional superconductor as: $\left(\frac{H_{c2}(\theta)\cos\theta}{H_{c2,\perp}}\right)^2 + \left(\frac{H_{c2}(\theta)\sin\theta}{H_{c2,\parallel}}\right)^2 = 1$. The conformity to the Tinkham formula is consistent with our observation of the (slight) violation of the Pauli limit. The $\theta$-dependent measurements thus suggest that the superconductivity in ZrAs$_2$ is in the two-dimensional limit.

To gain further insights into the superconducting transition, we conducted systematic temperature-dependent measurements Figures 2**d** and 2**e** provide a visual representation of the acquired results when the magnetic field was applied perpendicularly to the ZrAs$_2$ (001) plane. It is worth noting that the superconducting transition progressively shifts to lower magnetic fields with increasing temperature. This evolution of the temperature-dependent out-of-plane upper critical magnetic fields ($\mu_0 H_{c2,\perp}$) is summarized in Fig. 2**e**. Given that the angular dependence of $\mu_0 H_{c2}(\theta)$ can be described by a two-dimensional superconductivity model, it is anticipated that $\mu_0 H_{c2,\perp}$ would exhibit a linear relationship with temperature close to $T_c$ as given by the standard Ginzburg-Landau model for two-dimensional superconductors[54]: $\mu_0 H_{c2,\perp}(T) = \frac{\Phi_0}{2\pi\xi_{GL}(0)^2}\left(1 - \frac{T}{T_c}\right)$. Here $\xi_{GL}(0)$ stands for the zero-temperature in-plane coherence length, $\Phi_0$ is the magnetic flux quantum, and $T_c$ is the critical temperature. Indeed, near $T_c$, $\mu_0 H_{c2,\perp}$ displays linear behavior as a function of $T$ before dropping to zero at $T_c$ as expected for two-dimensional superconductivity. By fitting the data, we estimate $\xi_{GL}(0)$ to be 70 nm. We also examined the temperature-dependent superconducting transition with the magnetic field applied parallel to the sample $ab$ plane. Figure 2**f** show the corresponding traces, and Fig. 2**f** summarizes the result, illustrating the in-plane upper critical field $H_{c2,\parallel}$ as a function of temperature. Akin to $\mu_0 H_{c2,\perp}(T)$, $\mu_0 H_{c2,\parallel}(T)$ can be fitted to the standard Ginzburg-Landau model for two-dimensional superconductors[54]: $\mu_0 H_{c2,\parallel}(T) = \frac{\Phi_0\sqrt{3}}{\pi\xi_{GL}(0)d_{sc}}\left(1 - \frac{T}{T_c}\right)^{1/2}$ near $T_c$. From this fit, we extract a superconducting thickness $d_{SC} \approx 4.2$ nm, which is approximately 17 times smaller than $\xi_{GL}(0)$. Despite the good agreement with the fits to two-dimensional superconductivity, it should be noted that $d_{SC}$ is much smaller (by at least 4 orders of magnitude) than the sample thickness but at least 4 times larger than the $c$-axis lattice constant of ZrAs$_2$, suggesting that its superconductivity could be confined to just 4 unit-cells at ZrAs$_2$ (001) plane.

The apparent two-dimensional superconductivity in ZrAs$_2$ cannot be directly attributed to the dimensionality of the bulk Fermi surface, which we directly probed through our quantum oscillatory experiments as depicted in Extended Figs. 4 and 5. The magnetoresistance traces displayed in Extended Figs. 4 and 5 reveal distinct $1/\mu_0 H$-periodic quantum oscillations (or the Shubnikov-de Haas effect). By analyzing the quantum oscillatory data (detailed in Extended Figs. 4 and 5, with a discussion in Methods Section V), we observe that while the quantum oscillatory frequencies ($f_2$ and $f_3$) display dependence on magnetic field orientation, neither conforms to the $1/\cos\theta$ dependency characteristic of two-dimensional Fermi surfaces. This observation implies that the carriers arising from these bulk Fermi surfaces are unlikely to contribute to the Cooper pairing mechanism. As discussed below, we therefore argue that the driving factor behind the superconductivity in ZrAs$_2$ is associated with paired charge carriers on the surface state of the $ab$ plane.

Given that ZrAs$_2$ demonstrates a superconducting transition in resistivity measurements, one would anticipate the appearance of the Meissner effect. However, testing of multiple samples revealed that, despite the observable superconducting transition in the resistivity, the Meissner effect is absent down to 1.69 K (as shown in Extended Fig. 7). Furthermore, as detailed in Methods Section VII and Extended Fig. 8, our muon spin relaxation (μSR) measurements in ZrAs$_2$ show no signature of a bulk superconducting state down to 0.04 K. The most likely explanation is that the superconductivity is confined to the surfaces of ZrAs$_2$. Magnetic DC susceptibility and μSR measurements require a large sample volume for detection of the diamagnetic response associated with a superconducting transition. Consequently, if only the surface would superconduct, and not the bulk of the material, such a state will not be captured by these techniques. This distinction likely explains why we did not observe signatures for superconductivity in ZrAs$_2$ from the magnetic susceptibility or µSR measurements since the superconductivity in this material is limited to its surface with the non-superconducting bulk contribution dominating the magnetic susceptibility and the µSR signals. Notably, the angular dependence of the $\mu_0 H_{c2}(\theta)$ aligns with the possibility of superconductivity emerging solely in the *ab* plane of the orthorhombic crystal. Such a surface-confined superconducting state is expected to exhibit behavior characteristic of the two-dimensional limit, which is precisely what we observe in our measurements of $\mu_0 H_{c2}(\theta)$, $\mu_0 H_{c2,\perp}(T)$, and $\mu_0 H_{c2,\parallel}(T)$. It is also worth noting that a surface superconductor like ZrAs$_2$ is different from quasi-2D superconductors like YBa$_2$Cu$_3$O$_{7-\delta}$, which exhibit superconductivity in all layers, *i.e.*, in the bulk, therefore, its superconductivity can be detected through magnetic susceptibility and µSR measurements.

The observation of two-dimensional superconductivity in a bulk single crystal sample lacking a quasi-two-dimensional Fermi surface, coupled with the absence of bulk superconductivity, strongly suggests that the pairing mechanism is exclusively driven by carriers on the surface state. This unique surface superconductivity fundamentally differs from three-dimensional superconductors, as it should give rise to the spontaneous emergence of vortices due to the Berezinskii–Kosterlitz–Thouless (BKT) transition in two-dimensions[55,56]. Through an examination of the voltage (*V*)- current (*I*) relationship obtained via electrical transport measurements across various temperatures, we uncover clear signatures of a BKT transition that reinforce the notion of the surface-confined nature of the superconducting state. As illustrated by Fig. 3**a**, at low temperatures, distinctive nonlinearities manifest in the *V*(*I*) characteristics, conforming to a temperature-dependent power law[57]: $V \propto I^{a(T)}$, where $a(T) = 1 + \pi J_S(T)/T$, and $J_S$ represents the superfluid stiffness. The precise occurrence of the BKT transition is determined by the condition $\pi J_S(T_{BKT})/T_{BKT} = 2$, yielding $a(T_{BKT}) = 3$ and thereby defining the value of $T_{BKT}$. We identified such behavior in the *V*(*I*) characteristics, which are plotted in Figs. 3**a** and 3**b**, where a cubic power law is evident for $T \simeq 1.5$ K. Note that for the BKT transition in an infinite and homogeneous sample, a universal jump of $a(T)$ is anticipated at the BKT temperature ($T_{BKT}$), transitioning from $a(T > T_{BKT}) = 1$ to $a(T < T_{BKT}) = 3$. However, in practice all samples inherently possess inhomogeneities or defects and are of a finite size. Consequently, the discontinuity in $a(T)$ at $T_{BKT}$ is anticipated to broaden but must occur at a temperature between $T = T_c$ and $T = T_{BKT}$[57]. In ZrAs$_2$, we find that $a(T)$ decreases rather sharply with increasing temperature and falls to 1 within 0.08 K of $T_{BKT}$. Such a dramatic change in $a(T)$ is consistent with what one expects from the BKT transition.

For a consistency check, we also plot the superfluid stiffness $J_S(T)$ obtained from $\frac{T}{\pi}(a(T) - 1)$ in Fig. 3**c**. We find that at low temperatures, $J_S(T)$ saturates, and at elevated temperatures, $J_S(T)$ becomes smoothly suppressed as quasiparticles emerge when the temperature is raised, as qualitatively expected from the BCS theory; see Supplementary Information Section III and Supplementary Fig. 4 for the BCS fitting results. Importantly, above $T_{BKT}$, $J_S(T)$ drops much rapidly as one would expect from the BKT transition. Note that the power dissipated in the sample corresponding to the traces presented in Fig. 3**a** is maximum 50 picowatts. Such a small power dissipation is unlikely to cause Joule heating of the sample.

Next, we scrutinized the temperature dependence of the resistance near the superconducting transition. Using the Halperin–Nelson theory[58], Benfatto *et al.*[57] formulated the temperature dependence of the resistance for $T \geq T_{BKT}$ as $\frac{R(T)}{R_N} = \frac{1}{1+A_{HN}\sinh^2\left(\sqrt{B_{HN}\frac{T_c-T_{BKT}}{T-T_{BKT}}}\right)}$. Here, $R_N$ represents the normal state resistivity, $B_{HN}$ stands for a dimensionless constant of the order of 1, and $A_{HN}$ is a pre-factor. By fixing $T_{BKT}$ at the value obtained from the *V*(*I*) characteristics, namely $T_{BKT} = 1.5$ K, we achieve a good fit to the experimental data (Fig. 3**d**). The good agreement with the Halperin–Nelson formula underscores that the superconducting state in ZrAs$_2$ resembles that of a pure superconducting state, as the formula assumes a homogenous superconducting state with no broadening in $T_{BKT}$. This observation is indeed in harmony with the high quality of the sample, as indicated by its substantial residual-resistivity ratio and the clear presence of quantum oscillatory phenomena (see Methods Section V). Importantly, it is worth emphasizing that, the BKT transition has previously been observed only in quasi-2D-layered superconductors and thin films[59-63]. But to our knowledge, a BKT transition in a superconducting state confined solely to a specific surface of a three-dimensional crystal has not yet been reported. The observation of the BKT transition in a bulk ZrAs$_2$ crystal further accentuates the surface-confined nature of the superconducting state in ZrAs$_2$.

Surface superconductivity was previously claimed for CaAgP[64] and its Pd doped version as well as for PtBi$_2$[65]. For CaAgP and its Pd-doped variant, point contact spectroscopy would imply strong coupled, spin-triplet chiral superconducting pairing. However, such conclusions are at odds with the very small upper critical fields measured by the same authors, frequently seen in non-uniform, filamentary superconductors[66], thus suggesting that these hypotheses require additional experimental confirmation. As for PtBi$_2$, the conclusion in favor of superconductivity below $T_c \sim 10$ K is based on the loss of the spectral weight of the Fermi arcs. Nevertheless, relatively thick exfoliated flakes from this type-I Weyl compound, exhibit superconducting transitions well below 1 K, according to transport measurements[62]. Therefore, further work is required to elucidate the seemingly contradictory behavior of these superconductors. In contrast, our work on ZrAs$_2$ reveals a clear zero-resistance state, high planar upper critical fields, and a BKT-like superconducting transition, all in the absence of any evidence for bulk superconductivity, clearly suggesting a surface superconducting state.

The observation of surface-confined superconductivity in ZrAs$_2$ prompts us to explore the potential mechanisms that could stabilize such pairing in the surface state. To this end, we investigated the surface-projected Fermi surface and band structure. The *ab-initio* calculations (incorporating spin-orbit coupling) of the surface state projected onto the (001) plane, where we observe the superconducting state, shows termination-dependent surface state contribution. We consider the termination which matches our experimental data (see supplementary Information Section II and Supplementary Figs. 2 and 3). The calculations reveal a vHs located at the $\overline{X}$ point in the band corresponding to the surface state at approximately 65 meV below the Fermi level (see Fig. 4**a,b**). This saddle point is characterized by the crossing of the surface states at a constant energy contour (Fig. 4**c**). This observation is also evident in the three-dimensional view of the band structure of the surface state within the $k_x$-$k_y$ plane, around the $\overline{X}$ point, as depicted in Fig. 4**c**. The saddle point, or vHs, results from the intersection of two symmetric pockets in the constant energy contours at the high symmetry $\overline{X}$-point.

To experimentally probe the vHs around the $\overline{X}$ point, we conducted further ARPES measurements. Figure 4**d** illustrates the evolution of the constant energy contours across consecutive energy slices; the left panels display the experimental results, while the right panels depict the corresponding *ab-initio* calculations. As observed in Fig. 4**d**, two symmetry-related butterfly-wing-shaped states are separate at the Fermi level with the apex of each pocket lying along the $\overline{\Gamma} - \overline{X}$ direction. As the energy is lowered to -65 meV, the two states converge, ultimately leading to an intra-band intersection at the $\overline{X}$ point, thus forming a vHs. Further changes in energy towards -100 meV and -150 meV reveal butterfly-like features and their intersecting section near the $\overline{X}$ point becomes hole-like. The presence of vHs in the surface state is also evident in the energy-momentum slices along the $\overline{\Gamma} - \overline{X} - \overline{\Gamma}$ and $\overline{S} - \overline{X} - \overline{S}$ directions. Figure 4**e** illustrates these results, showcasing the energy-momentum dispersion spectra on the (001) surface projected *ab-initio* calculations. Notably, the (001) surface state displays an electron-like dispersion along the $\overline{\Gamma} - \overline{X} - \overline{\Gamma}$ direction and a hole-like dispersion along the $\overline{S} - \overline{X} - \overline{S}$ direction, thus highlighting the formation of a saddle point at the $\overline{X}$ point. Therefore, our ARPES spectra combined with *ab-initio* calculations unveil a vHs around the $\overline{X}$ point in the electronic surface bands of the (001) plane, precisely where we observe the two-dimensional superconducting state.

To further distinguishing between surface and bulk contributions to the observed vHs, we performed a detailed photon-energy-dependent ARPES study. Figure 5 summarizes our results. Our analysis reveals that at the binding energy $E_\text{B} = E_\text{vHs}$, the contour of the vHs state near the $\overline{\text{X}}$ point remains consistent across the photon energy range studied (35–55 eV), as illustrated in Fig. 5f. Furthermore, the peaks in the momentum distribution curves at $E_\text{B} = E_\text{vHs} - 0.1$ eV, corresponding to the hole dispersion of the VHS state along the $\overline{\text{S}}$-$\overline{\text{X}}$-$\overline{\text{S}}$ direction, show no variation with photon energy, as illustrated in Fig. 5f. In contrast, the bulk state near the $\overline{\Gamma}$ point shows significant variation with photon energy, underscoring the distinct behavior inherent to bulk states. This absence of photon energy dependence for the vHs state suggests that it results from a surface state, as surface states are independent of photon energy, unlike bulk states. We refer the readers Extended Fig. 9a-c for additional results supporting that the vHs is formed by the surface state.

The presence of a vHs in the $k_x$-$k_y$-plane within the first Brillouin zone indicates an enhanced density of states at -65 meV, which is slightly below the Fermi energy (see Extended Figs. 10 for the ARPES results and 11 for the *ab-initio* calculations) within that plane. Extended Fig. 10 presents energy distribution curves along two orthogonal directions passing through the $\overline{\text{X}}$ point. We find that the peak observed at the top band along the $\overline{S} - \overline{X} - \overline{S}$ direction aligns with the bottom of the band along the $\overline{\Gamma} - \overline{X} - \overline{\Gamma}$ path, forming a two-dimensional saddle point. This saddle point in the surface state dispersion naturally results in a vHs, owing to the inevitable divergence of the density of states at the saddle point of a two-dimensional electronic state[67,68]. The corresponding increased density of states, which can be further explored via future tunneling measurements, has the potential to induce electronic order, such as superconductivity. Our transport experiments, presented in Figs. 2 and 3, reveal that the superconducting state is confined within the *ab*-plane, which is directly connected with the $k_x$-$k_y$-plane in the first Brillouin zone where the vHs is located (Fig. 4). Hence, it is likely that the vHs, and concomitant enhanced density of states, play a pivotal role in stabilizing the superconducting state on the same surface.

Our discovery carries several implications: first, it establishes a convenient material platform for investigating superconductivity on the surfaces of a topological material, with this being important for the field of topological superconductivity[12,69]. Second, it provides a material foundation for exploring the interplay between vHs and superconductivity at low dimensions (see our theory discussion in Methods Section VIII and Supplementary Information Section IV for possible pairing symmetries for the nontrivial topology of the superconducting state), paving the way for the discovery of unconventional quasiparticles such as Majorana fermions. The unprecedented surface-confined nature of both the vHs and superconductivity opens avenues for studying low-dimensional physics, akin to the correlation-induced low-dimensional superconductivity observed in twisted bilayer graphene, but in bulk materials. Third, it provides a strategy, beyond the proximity effect, for observing potentially unconventional superconductivity at the surface of topologically non-trivial compounds. Finally, given that these quantum phenomena emerge at the surface of a topological material, they are expected to exhibit robustness against local perturbations, thus exhibiting potential for applications in quantum devices that harness surface-confined superconductivity.

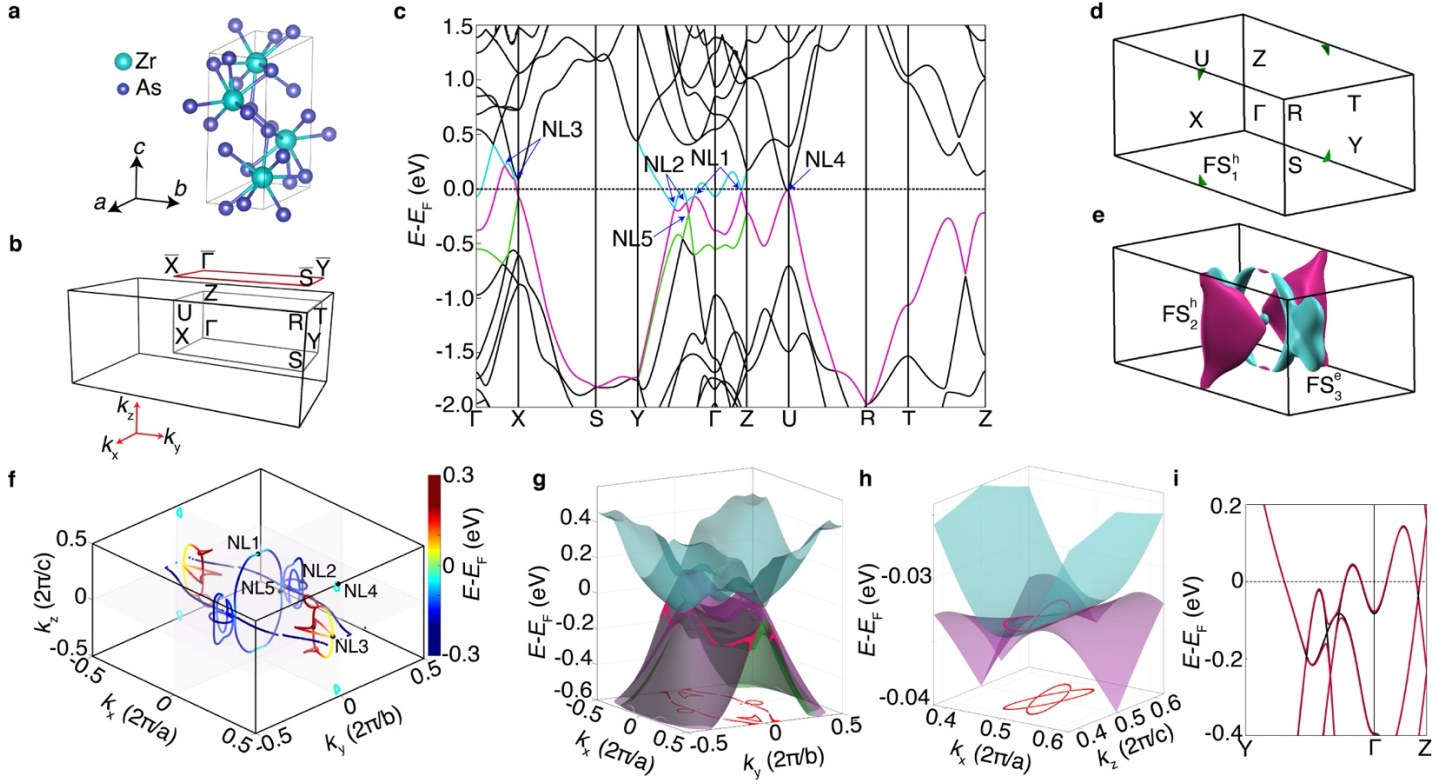

**Fig. 1: Electronic band structure of ZrAs$_2$ displaying nodal lines. a**, Crystal structure of ZrAs$_2$. **b**, First Brillouin zone (surface and bulk) of ZrAs$_2$, highlighting its high symmetry points and the projection of the (001) surface. **c**, Electronic band structure of ZrAs$_2$ obtained using first-principles calculations (without considering spin-orbit coupling) along high-symmetry lines. **d**, **e**, Three-dimensional Fermi surface of ZrAs$_2$ displaying two hole-like (FS$_1^h$ and FS$_2^h$) and one electron-like (FS$_3^e$) pockets. Color code corresponds to the bands shown in panel **c**. **f**, Nodal loops in the $k_x$-$k_y$-$k_z$ space formed by the band crossings between green, magenta, and cyan bands in panels **c**, **g**, **h**. Three-dimensional visualization of the electronic band crossings and the nodal lines or loops (red lines) within the $k_z = 0$ (panel **g**) and $k_y = 0$ planes near the U point (panel **h**). Bands are shown using the same color as in **c**. Projection in $E$-$E_F = 0$ plane indicates the location of the nodes in momentum space. **i,** Enlarged view of the band structure along the Y → Γ → Z direction, contrasting bands with (red) and without (black) spin-orbit coupling.

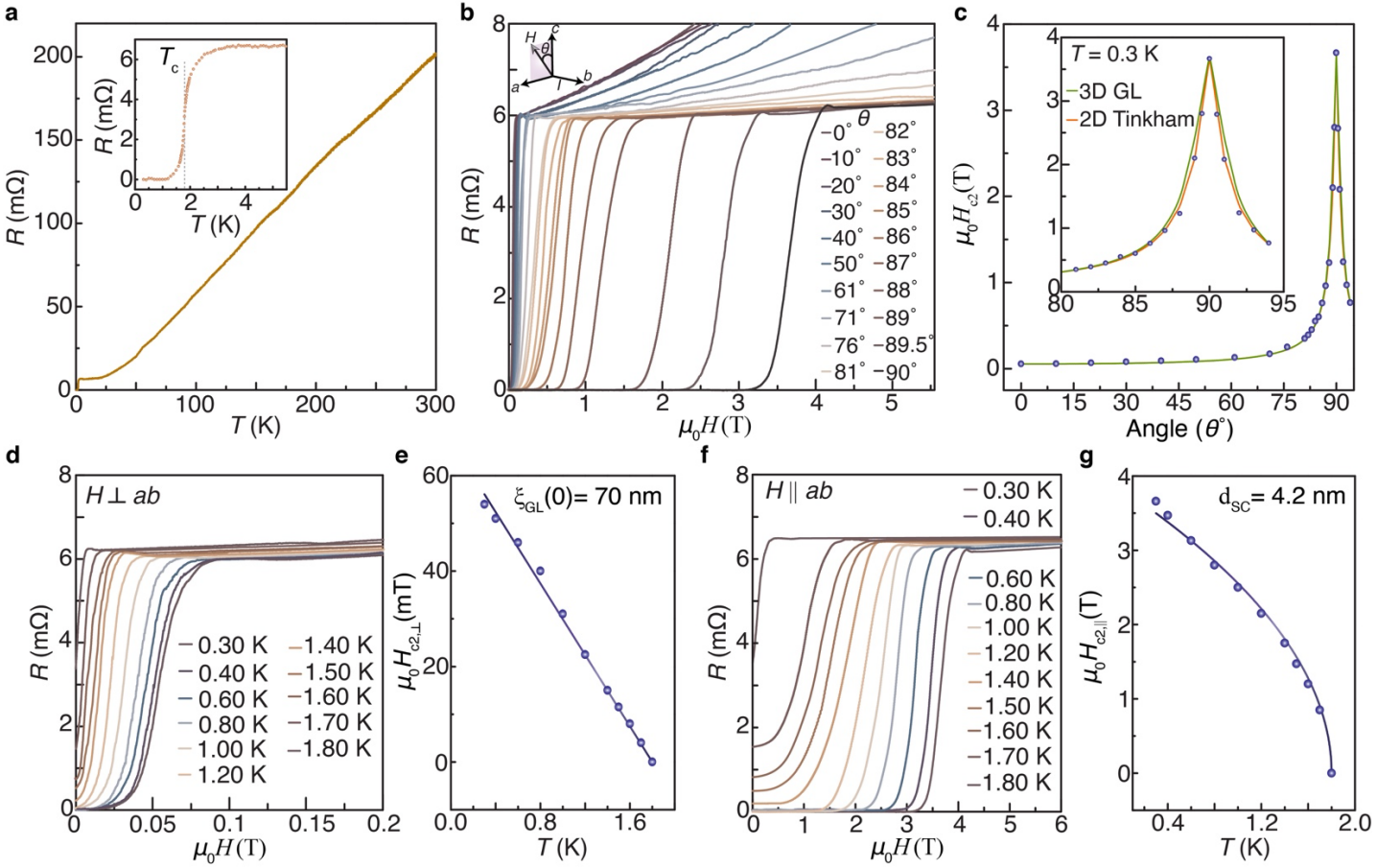

**Fig. 2: Superconducting instability in ZrAs$_2$. a,** Four-probe resistance ($R$) of a small ZrAs$_2$ single-crystal, exhibiting metallic behavior as a function of the temperature ($T$) before transitioning to a superconducting state at $T_c \simeq 1.8$ K. The inset depicts a magnified image of the resistance as a function of the temperature trace near $T_c$. **b,** Resistance at $T \simeq 0.3$ K as a function of the magnetic field for different out-of-the-plane rotation angles ($\theta$). The inset illustrates the direction of rotation, where $\theta$ represents the angle between the out-of-plane magnetic field and the $c$-axis of the crystal. **c,** $\theta$-dependence of the upper critical magnetic field ($\mu_0 H_{c2}$), defined as the field at which the resistance becomes half of its value in the normal state just above the transition. The orange and green curves represent the values expected from the Tinkham formula for a two-dimensional superconductor and the three-dimensional Ginzburg-Landau anisotropic mass model, respectively, considering the observed out-of-plane upper critical field ($\mu_0 H_{c2,\perp} = 0.056$ T) and in-plane upper critical field ($\mu_0 H_{c2,\parallel} = 3.66$ T). Inset: A magnified view of the same data near $\theta = 90°$. The data show excellent agreement with the Tinkham formula while deviating from the three-dimensional Ginzburg-Landau anisotropic mass model, particularly near $\theta = 90°$. **d,** Resistance as a function of the out-of-plane magnetic field for different temperatures. **e,** Temperature dependence of $\mu_0 H_{c2,\perp}$, displaying a linear dependence on $T$ close to $T_c$. Dark blue line represents a fit to the Ginzburg-Landau model for two-dimensional superconductors, yielding an in-plane coherence length of $\xi_{GL}(0) \simeq 70$ nm. **f,** Resistance as a function of the out-of-plane magnetic field for different temperatures. **g,** Temperature dependence of $\mu_0 H_{c2,\parallel}$. Dark blue line depicts a fit to the Ginzburg-Landau model for two-dimensional superconductors, yielding a superconducting thickness of $d_{SC} \simeq 4.2$ nm.

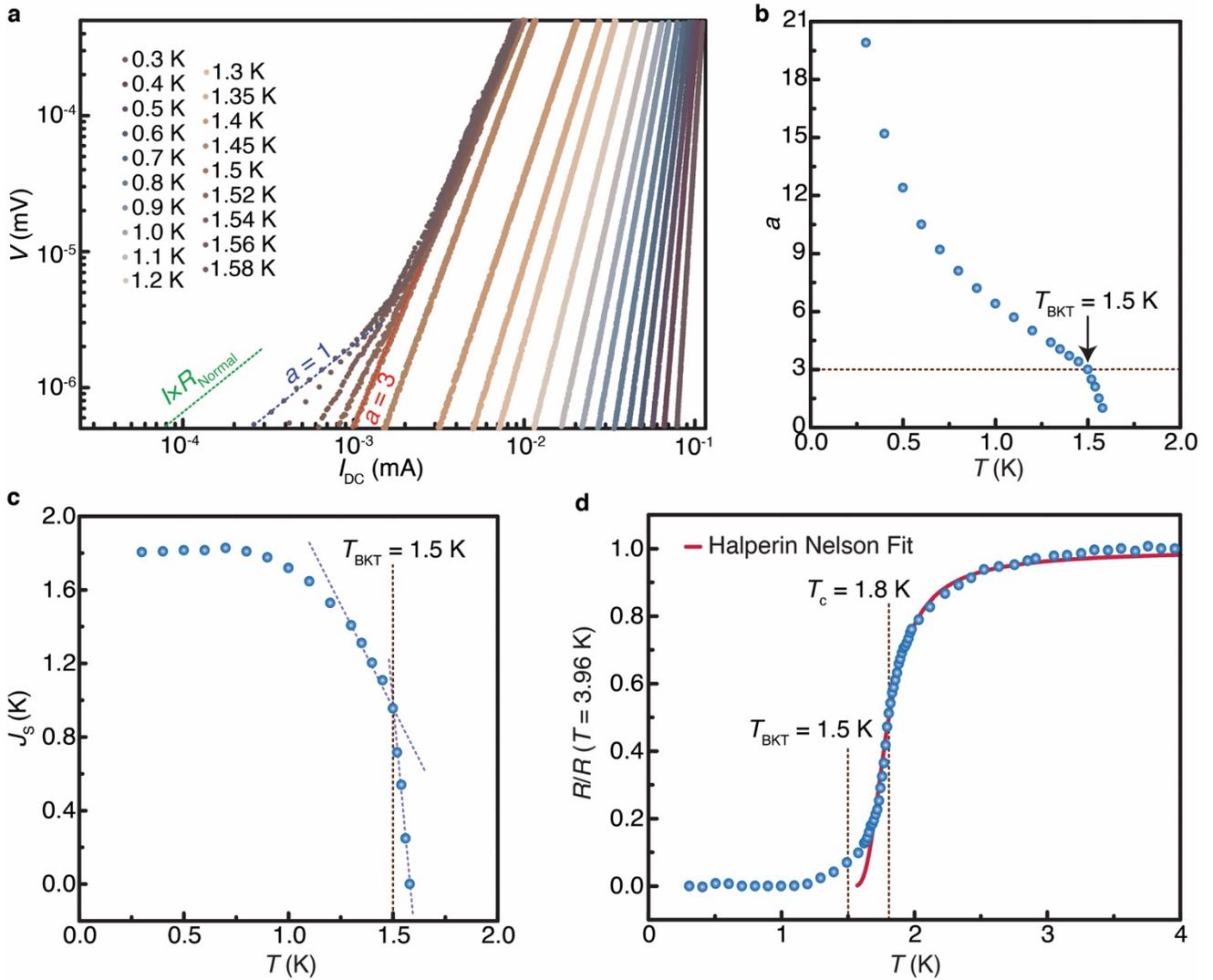

**Fig. 3: Observation of the Berezinskii–Kosterlitz–Thouless (BKT) transition. a**, Voltage ($V$)-current ($I$) characteristics acquired in the low-current regime, in a logarithmic scale, and for different temperatures. The red, blue, and green dashed lines represent $V \propto I^3$, $V \propto I$, and $V = I R_{\text{Normal}}$, respectively. **b**, Temperature dependence of the exponent $a(T) = 1 + \pi J_S(T)/T$ extracted from fitting the $V$-$I$ data to $V \propto I^{a(T)}$. The BKT transition is characterized by $\pi J_S(T_{\text{BKT}})/T_{\text{BKT}} = 2$, leading to $a(T_{\text{BKT}}) = 3$, the relation that defines $T_{\text{BKT}}$. The horizontal dashed line indicates the value $a = 3$. $a(T)$ as a function of $T$ clearly reveals the BKT transition with $T_{\text{BKT}} \simeq 1.5$ K. **c**, Temperature dependence of the superfluid stiffness, $J_S$. $J_S$ drops near $T_c$, as expected from the BCS theory. Above $T_{\text{BKT}} \simeq 1.5$ K, the $J_S$ decreases more rapidly with increasing $T$, as shown by the violet dashed lines serving as guides to the eye. **d**, Temperature dependence of the resistance (blue circles), fitted to the Halperin–Nelson theory using the parameters $T_{\text{BKT}} = 1.5$ K and $T_c = 1.8$ K. Red curve represents the best fit.

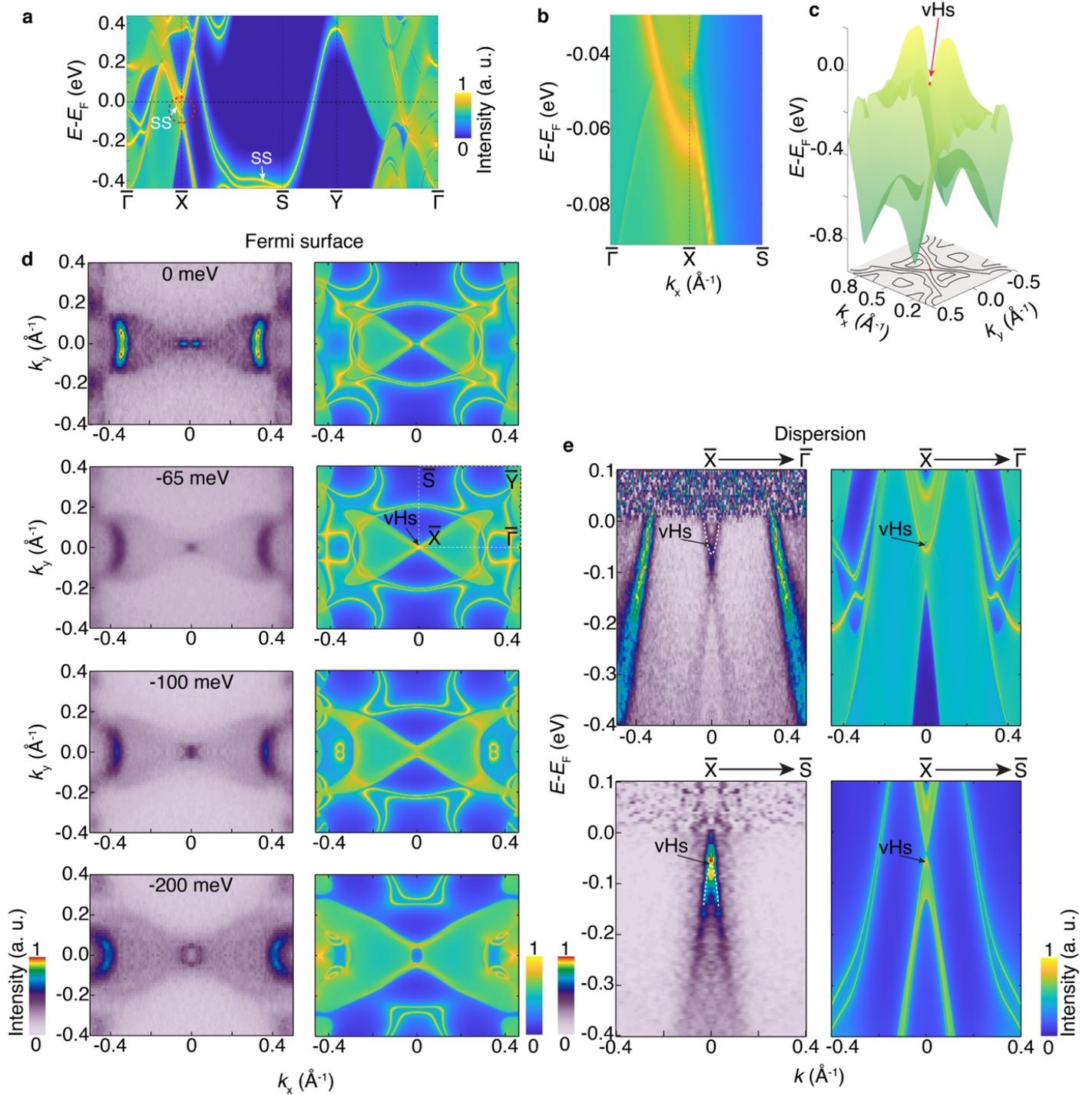

**Fig. 4: Surface van Hove singularity near the Fermi energy. a**, *Ab-initio* calculations (considering spin-orbit coupling) of the band structure projected onto the (001) surface, revealing the presence of surface states (SS), marked with arrows, and a van Hove singularity (vHs) at the $\bar{X}$ point. **b**, Magnified view of the saddle point at the $\bar{X}$ point, marked with a red circle in panel **a**. **c**, Three-dimensional view of the surface state dispersion near the vHs at the $\bar{X}$ point on the $k_x$-$k_y$ plane with $k_z$ fixed at 0 Å. **d**, Angle-resolved photoemission spectra (left column) and corresponding *ab-initio* calculations (considering spin-orbit coupling) illustrating the surface-projected constant energy contours at $E_b$ = 0 eV, -65 meV (where the vHs is located), -100 meV, and 200 meV. For convenience, the coordinates of the $\bar{X}$ point are chosen to correspond to the (0, 0) coordinates. The photoemission spectroscopy results align with the calculated constant energy contours. The black arrow on the Fermi surface of the surface state indicates the location of the vHs, identified by the intersections among surface states. **e**, Energy-momentum slices along the $\bar{\Gamma} - \bar{X} - \bar{\Gamma}$ and $\bar{S} - \bar{X} - \bar{S}$ directions acquired with photoemission spectroscopy (left column) and the (001) surface-projected calculation (right). The surface states (marked by white dashed parabolic bands) exhibit electron-like dispersion along $\bar{\Gamma} - \bar{X} - \bar{\Gamma}$ and hole-like dispersion along $\bar{S} - \bar{X} - \bar{S}$ direction, forming a saddle point (marked by the black arrow) at the $\bar{X}$ point. For each binding energy, the energy-momentum cuts (panel **e**) are normalized at each binding energy to highlight band intensities near the Fermi level.

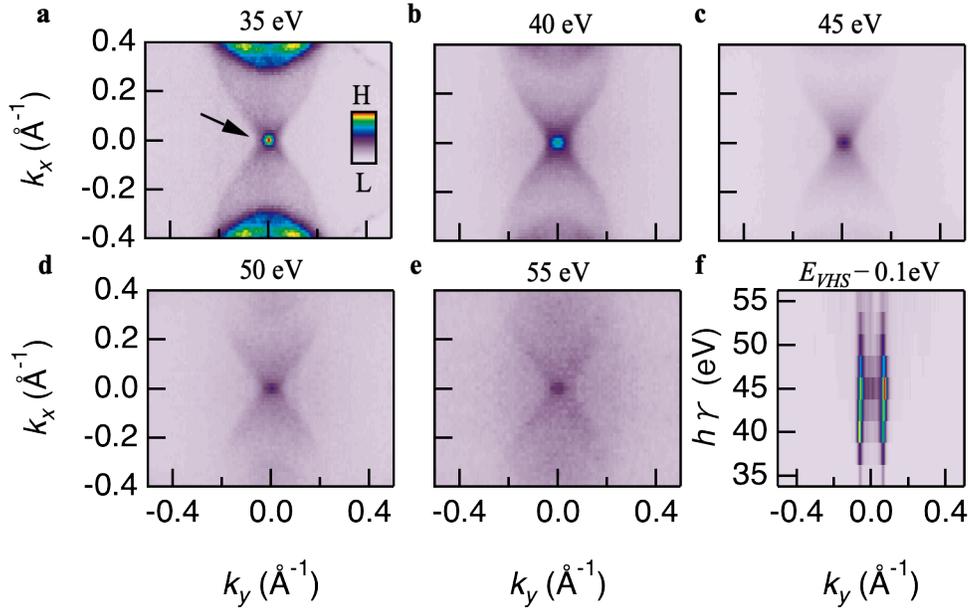

**Fig. 5. Photon energy dependence of the van Hove singularity (vHs) state at the $\bar{X}$ point. a-e**, Constant energy contours at the fixed energy of the vHs ($E_{vHs}$), measured at photon energies of 35, 40, 45, 50, and 55 eV, respectively. Note that the maps are symmetrized with respect to $k_x = 0$. **f**, Photon energy dependence of the momentum distribution curves measured at $E_{vHs} - 0.1 eV$ along the $\bar{S}$-$\bar{X}$-$\bar{S}$ direction. The contour of the vHs state remains unchanged across different photon energies, demonstrating its surface state nature.

**Methods:**

### I. Synthesis of ZrAs₂ single crystals:

ZrAs$_2$ single crystals were synthesized through a two-step chemical vapor transport (CVT) technique[70]. High-purity precursors, consisting of Zirconium (Zr) with a purity of 99.98% and Arsenic (As) foil with a purity of 99.99%, were combined in a stoichiometric ratio of 1:2. The mixture was introduced along with iodine as a transport agent into quartz tubes under a vacuum pressure of $5 \times 10^{-4}$ Torr. The quartz tubes were subsequently placed in a two-temperature commercial CVT setup, maintaining the lower and higher temperatures at 850°C and 750°C, respectively, for a continuous five-day period. The reaction was then gradually cooled down to room temperature at a rate of 100°C/h, resulting in the formation of polycrystalline samples. The obtained polycrystals were finely ground and then subjected to the same CVT process using iodine as the transport agent. This process was conducted with the same temperature gradient for a duration of two weeks.

Following this, needle-like single crystals of typical dimensions ranging from 0.5 to 1 mm (in length) were successfully obtained in the hotter end (Extended Fig. 1**a**). The crystallographic orientation and structure of the single crystals were determined using Powder X-ray diffraction (XRD) measurements (Extended Fig. 1**b**). The XRD patterns were matched with the DB Card No: 01-079-6540 (ICDD (PDF-4+ 2020 RDB)). The XRD data further confirmed the Cotunnite mineral structure of ZrAs$_2$, crystallizing in the orthorhombic Pnma (62) space group with lattice parameters $a = 6.8016(10)$ Å, $b = 3.6889(5)$ Å, and $c = 9.0305(13)$ Å.

We have also performed energy-dispersive X-ray spectroscopy (EDS) measurements on the ZrAs$_2$ crystals to determine the exact stoichiometric ratio of Zr and As in the grown crystals. The results are summarized in Extended Figs. 2 and 3, and Extended Tables 1 and 2. Various portions of the crystal, shown in Extended Fig. 2**a**, were examined to estimate the chemical composition. The selected area revealed that, except for carbon (used to adhere the sample to the holder), only Zr and As peaks are visible in the EDS spectrum. We further tested different points and areas to determine the stoichiometric ratio of the elements in the compound, as detailed in Extended Tables 1 and 2. Extended Table 1 presents the specific ratios found in different parts, while Extended Table 2 displays the calculated mean values derived from Extended Table 1. These results unequivocally confirm the exact stoichiometric ratio of Zr to As as 1:2.

## II. First-principles calculations

The electronic structure calculations were conducted using the projector augmented wave method within the density functional theory framework, employing the VASP package[71]. A plane-wave energy cut-off of 650 eV was utilized for the study. The generalized gradient approximation (GGA) method served as the exchange-correlation functional[72]. Electronic band structure calculations were performed using a 6 × 12 × 4 Monkhorst-Pack $k$-mesh, with self-consistent incorporation of spin-orbit coupling. The VASPWANNIER90 interface was utilized, utilizing $s$ and $d$-orbitals of Zr atoms and $p$-orbitals of As atoms, to generate a tight-binding Hamiltonian without employing the procedure for maximizing localization[73,74]. The calculation of the surface state was performed using the semi-infinite Green's function approach integrated into the WannierTools[75,76]. We performed all the calculations with spin-orbit coupling except for the data shown in Fig.1**c**-1**h**.

## III. Angle-resolved photoemission spectroscopy measurements

The angle-resolved photoemission spectroscopy measurements were performed at the beamline BL21_ID (ESM) of National Synchrotron Light Source II (NSLS II). The samples were cleaved under ultra-high vacuum conditions and measured at $T = 10$ K. The data were obtained using a photon energy of 40 eV with $p$ polarization. The energy and momentum resolutions are better than 15 meV and 0.01 Å$^{-1}$, respectively. Due to the strong interatomic bonds in ZrAs$_2$, cleaving this material is particularly challenging, and the resulting flat terraces that yield sharp ARPES spectra are typically around 100 × 100 μm$^2$ in size. To overcome this challenge, we cleaved multiple samples and utilized a small beam size ($< 30 \times 50$ μm$^2$) at BL21 to investigate the band structure within a single flat region.

## IV. Transport measurements:

Low-field transport measurements (presented in Figs. 2 and 3) were conducted using an Oxford Heliox system, which has a base temperature of 0.3 K and a maximum magnetic field capability of 8 T. To obtain the temperature-dependent quantum oscillations data (presented in Extended Fig. 4), we used the WM5 water-cooled magnet generating magnetic fields up to 35 T at the Steady High Magnetic Field Facilities of the High Magnetic Field Laboratory, Chinese Academy of Sciences in Hefei, China. On the other hand, for angle-dependent magnetoresistance measurements acquired at $T = 5$ K (presented in Extended Fig. 5), we utilized a standard 4He cryostat and magnetic fields up to approximately 28.3 T.

## V. Quantum oscillations via transport measurements:

Quantum oscillation experiments provide crucial insights into the fermiology of quantum materials. Here, we investigated the fermiology of ZrAs$_2$ by conducting magnetoresistance measurements and leveraging the Shubnikov- de Haas effect.

In Extended Fig. 4**a**, we show the magnetoresistance traces acquired at different temperatures. After removing background signals, the data unveils distinct quantum Shubnikov- de Haas oscillatory patterns (Extended Fig. 4**b**) corresponding to different temperatures. Utilizing the fast Fourier transform (FFT) analysis, we identify three oscillation peaks, denoted as $f_1$, $f_2$, and $f_3$ within the FFT spectra (Extended Fig. 4**c**), signifying the presence of three distinct Fermi pockets. We can obtain carrier effective masses associated with these three Fourier transform peaks by measuring the temperature dependence of their magnitudes. Extended Fig. 4**b-d** summarizes such data. As the temperature is raised, the oscillation amplitudes gradually diminish (Extended Fig. 4**b**), leading to weaker peaks in the Fourier transform (Extended Fig. 4**c**). In Extended Fig. 4**d**, we illustrate the magnitude of the Fourier transform as a function of temperature. To extract the effective masses ($m^*$) associated with the three peaks, we fit their temperature-dependent data using the well-known Lifshitz–Kosevich temperature damping formula[77] $R_T \propto T/sinh(m^*T/B)$. The deduced carrier effective masses are as follows: 0.21 $m_e$ ($f_1$), 0.26 $m_e$ ($f_2$), and 0.31 $m_e$ ($f_3$) for the three distinct Fermi pockets.

Next, we performed angular-dependent magnetoresistance measurements at a temperature of 5 K, where the magnetic field was systematically tilted in different directions characterized by rotational angles. Specifically, the magnetic field orientation was adjusted while maintaining the same current path, ranging from the *c*-axis [001] ($\theta = 0^o$) to the [100] direction ($\theta = 90^o$) during the experiments. The analysis of resistance against the magnetic field data reveals a pronounced angle-dependent behavior in the magnetoresistance as the rotational angles change, spanning from $\theta = 0^o$ to $90^o$, as presented in Extended Fig. 5**a**. After removing background signals, the data unveils distinct quantum Shubnikov- de Haas oscillatory patterns (Extended Fig. 5**b**) corresponding to different magnetic field directions. FFT analysis yields three angle-dependent oscillation peaks, denoted as $f_1$, $f_2$, and $f_3$ within the FFT spectra (Extended Fig. 5**c**). By tracing the frequency positions of these peaks linked to the Fermi surface, we derive the angular dependence of the peak positions, as depicted in Extended Fig. 5**c**. The positions of these peaks shift as the angle changes indicating an anisotropic bulk Fermi surface. The $f_2$ peak positions in the FFT spectra exhibit an increasing frequency trend with rotational angles up to 40º, followed by a gradual decrease. Conversely, the $f_3$ peak positions exhibit a gradual increase with $\theta$. It is worth noting that the $f_1$ peak is located at a relatively low frequency, making it challenging to precisely track its angular dependence. The polar plot in Extended Fig. 5**d** illustrates the angular dependence of the $f_2$ and $f_3$ peaks, both highlighting the anisotropic nature of the Fermi surfaces. It is important to emphasize that none of the peaks demonstrate the characteristic $1/cos\theta$ dependence typical of two-dimensional Fermi surfaces. The absence of such behavior indicates that none of the bulk Fermi pockets connected to the observed quantum oscillation peaks possess a quasi-two-dimensional character. This leads to the reasonable inference that the carriers arising from these bulk Fermi surfaces play a minimal role on the Cooper pairing mechanism within the observed two-dimensional superconducting state. Consequently, the emergence of two-dimensional superconductivity in ZrAs$_2$ is more plausibly linked to the Cooper pairing of charge carriers within the surface state.

### VI. Magnetic susceptibility measurements

Magnetic susceptibility measurements were conducted using a DynaCool Quantum Design Physical Properties Measurement System (PPMS) from Quantum Design (USA), equipped with a 9 T magnetic field. We acquired moment as a function of the temperature (*M-T*) curves employing two modes: zero field-cooled (ZFC), where data collection occurred during warming under an applied field after cooling under zero field, and field-cooling (FC), during which data was similarly collected while warming. An applied magnetic field (*H*) of 50 Oe was maintained, aligned parallel to the *ab* plane (*H* ∥ *ab*).

The magnetic susceptibility measurements were performed down to $T = 1.69$ K, the temperature limit of our magnetic susceptibility system. This temperature is reasonably below the superconducting transition temperature, $T_c$ =1.8 K.

Throughout the measurements, the sample exhibited a slight temperature-dependent variation after several runs. To ensure accuracy, we opted to take 10 data points for each temperature increment and subsequently calculated the average value. Extended Fig. 7 showcases our magnetic susceptibility data plotted against temperature for both ZFC and FC conditions. The magnetic susceptibility of the $ZrAs_2$ sample unveils a diamagnetic response across the entire temperature range for both ZFC and FC conditions. We observed no superconducting transitions in magnetic susceptibility measurements, although there was a slight increase in the diamagnetic signal at lower temperatures in the FC curve compared to the ZFC. This observation suggests that $ZrAs_2$ does not display the Meissner effect, indicating that the observed superconductivity in $ZrAs_2$ is not related to bulk superconductivity. It is worth emphasizing that magnetic susceptibility measurements primarily reflect the bulk properties of the sample, since it lacks sensitivity to detect the diamagnetic signal inherent to a superconducting surface. Since our electrical transport measurements, which are sensitive to both surface and bulk electronic states, indeed detect superconductivity in $ZrAs_2$, while the magnetic susceptibility technique, being fundamentally attuned to bulk characteristics, does not, we can conclude that the superconductivity in $ZrAs_2$ is localized exclusively on the surface of the compound.

### VII. Muon spin relaxation measurements

This section presents muon spin relaxation (μSR) data on $ZrAs_2$ obtained using the GPS instrument and the high-field HAL-9500 instrument, which is equipped with a BlueFors vacuum-loaded cryogen-free dilution refrigerator. These measurements were conducted at the Swiss Muon Source (SμS) at the Paul Scherrer Institut in Villigen, Switzerland. In Extended Fig. 8**a**, we present the transverse-field μSR-time spectra for $ZrAs_2$, measured under an applied magnetic field of $\mu_0 H = 5$ mT. The spectra above (2.2 K) and below (0.04 K) the superconducting transition temperature ($T_c$) are shown. The transverse-field-μSR data, indicated by solid lines in Extended Fig. 8**a**, were analyzed using the following functional form: $A_{TF_S}(t) = A_S e^{\left[-\frac{(\sigma_{sc}^2 + \sigma_{nm}^2)t^2}{2}\right]} \cos(\gamma_\mu B_{int} t + \phi)$. Here, $A_S$ denotes the initial asymmetry, $\frac{\gamma_\mu}{2\pi} \simeq 135.5$ MHz/T is the gyromagnetic ratio, and $\phi$ is the initial phase of the muon-spin ensemble. $B_{int}$ represents the internal magnetic field at the muon site, and $\sigma = \sqrt{\sigma_{sc}^2 + \sigma_{nm}^2}$ is the muon spin relaxation rate, where $\sigma_{sc}$ and $\sigma_{nm}$ characterize the damping due to the formation of the flux-line lattice in the superconducting state and the nuclear magnetic dipolar contribution, respectively. The transverse-field-spectrum in the normal and superconducting states almost perfectly overlaps, and the extracted depolarization rate, depicted in Extended Fig. 8**b** for two sets of positron detectors, shows no increase across $T_c$ or even at lower temperatures. Throughout the entire temperature range, the oscillations exhibit only a small relaxation due to the random local fields from the nuclear magnetic moments, with no additional contribution from the superconducting state below $T_c$. If the sample were a bulk superconductor, we would expect to see a clear increase in the depolarization rate below $T_c$ due to the presence of a non-uniform local magnetic field distribution resulting from the formation of a flux-line lattice in the superconducting state. However, this is not observed for $ZrAs_2$. This lack of increase in the relaxation rate below $T_c$ suggests that the system is not a bulk superconductor. Alternatively, if the sample was a bulk superconductor but had a very long penetration depth (more than 1000 nm), we would also not observe an increase in the muon spin depolarization rate below $T_c$. According to transport measurements, the penetration depth is relatively short (the superconducting thickness being only 4.2 nm, is equivalent to approximately four unit cells). Therefore, we conclude that the absence of an increase in the relaxation rate below $T_c$ is due to the absence of bulk superconductivity.

We also performed measurements under zero-field cooled conditions (*i.e.*, applying the field once in the superconducting state) and found no increase in the relaxation rate, which further suggests that the system does not exhibit bulk superconductivity. Thus, we exclude the long penetration depth as the reason for the temperature-independent rate across $T_c$. Even in the case of a long penetration depth, zero-field cooled measurements should result in a measurable relaxation rate within the superconducting state, which is not observed here.

## VIII. Theoretical discussion on the pairing symmetry of the superconducting state in ZrAs$_2$

To characterize the superconducting pairing symmetries on the surface of ZrAs$_2$, we consider the spin-split electronic states in the normal state due to the broken inversion symmetry near the surface, except at the time-reversal invariant momenta in the surface Brillouin zone. Our *ab-initio* calculations reveal that the spin-orbit coupling notably exceeds the experimentally obtained $k_B T_c$ of the superconducting state. Therefore, the Cooper paired state should comprise both spin singlet and triplet states[78]. Let $|k\rangle = \hat{\psi}_k^\dagger |0\rangle$ be the Bloch energy eigenstate of a surface spin-orbit coupled Hamiltonian with energy $\varepsilon_k$ at momentum $k$. Here, $\hat{\psi}_k^\dagger$ represents the electron quasiparticle creation operator in momentum space, satisfying the usual anticommutation relation $\{\hat{\psi}_k^\dagger, \hat{\psi}_{k'}\} = \delta_{kk'}$. Under time reversal operation, $\Theta$, the Bloch states transform as $|\tilde{k}\rangle = \Theta|k\rangle = \chi_k|-k\rangle$. Given $\Theta^2 = -1$, the phase factor $\chi_k$ satisfies $\chi_k^* \chi_{-k} = -1$[79]. For field operators, this transformation implies that, $\hat{\tilde{\psi}}_k^\dagger = \Theta \hat{\psi}_k^\dagger \Theta^{-1} = \chi_k \hat{\psi}_{-k}^\dagger$ and $\hat{\tilde{\psi}}_k = \Theta \hat{\psi}_k \Theta^{-1} = \chi_k^* \hat{\psi}_{-k}$. Superconducting pairing occurs between electrons states at $|k\rangle$ and its time-reversal partner $|\tilde{k}\rangle$. The condensate is described by $b_k = \langle \hat{\tilde{\psi}}_k^\dagger \hat{\psi}_k^\dagger \rangle$. Under the momentum reversal $k \to -k$, the condensate remains even, evident from $b_{-k} = \langle \hat{\tilde{\psi}}_{-k}^\dagger \hat{\psi}_{-k}^\dagger \rangle = -\chi_k^* \chi_{-k} \langle \hat{\tilde{\psi}}_k^\dagger \hat{\psi}_k^\dagger \rangle = b_k$.

The evenness of the superconducting condensate implies that the pairing symmetry should be an even function of momentum as well. Therefore, the superconducting gap functions transform under those irreducible representations whose basis functions are even. The surface of ZrAs$_2$ possesses low symmetry. Its symmetry group consists of only two elements, the identity and mirror reflection, $G_y = \{E, M_y\}$, whose actions are as $E: (x, y) \to (x, y)$ and $M_y: (x, y) \to (x, -y)$. This simple group is equivalent to the dihedral group, $D_1$, in 2D, which admits two one-dimensional irreducible representations transforming symmetrically and antisymmetrically with respect to the reflection operator $M_y$. For instance, the lowest order symmetric basis functions include $s_1$, $p_x \equiv k_x$, and $d_{x^2-y^2} \equiv k_x^2 - k_y^2$, while the antisymmetric ones are $p_y \equiv k_y$ and $d_{xy} = k_x k_y$ in momentum space.

As explained above, the *p*-wave pairings, due to the oddness of $p_x$ and $p_y$, are not allowed by symmetry. Therefore, the superconducting gap function may exhibit *s*-wave or *d*-wave symmetry. While isotropic *s*-wave symmetry complies with symmetry constraints, the surface superconducting states in ZrAs$_2$ could manifest anisotropic pairing symmetries such as $d_{x^2-y^2} \equiv k_x^2 - k_y^2$ or $d_{xy} = k_x k_y$. The latter pairings have nodes on the Fermi surface. However, the complex pairings $d_{x^2-y^2} \pm id_{xy}$ yield fully gapped superconducting states and are energetically more favorable thanks to enhanced condensation energy. The superconducting state with $d_{x^2-y^2} \pm id_{xy}$ pairing symmetry breaks the time-reversal symmetry, rendering it topologically nontrivial, *i.e.*, characterized by $\vartheta = \pm 2$ Chern number. The latter chiral superconducting states can potentially be explored in future magneto-optical Kerr measurements [80,81].

**Methods only reference**

70. Namba, R. et al. Two-step growth of high-quality single crystals of the Kitaev magnet $\alpha$−RuCl$_3$. *Phys. Rev.*

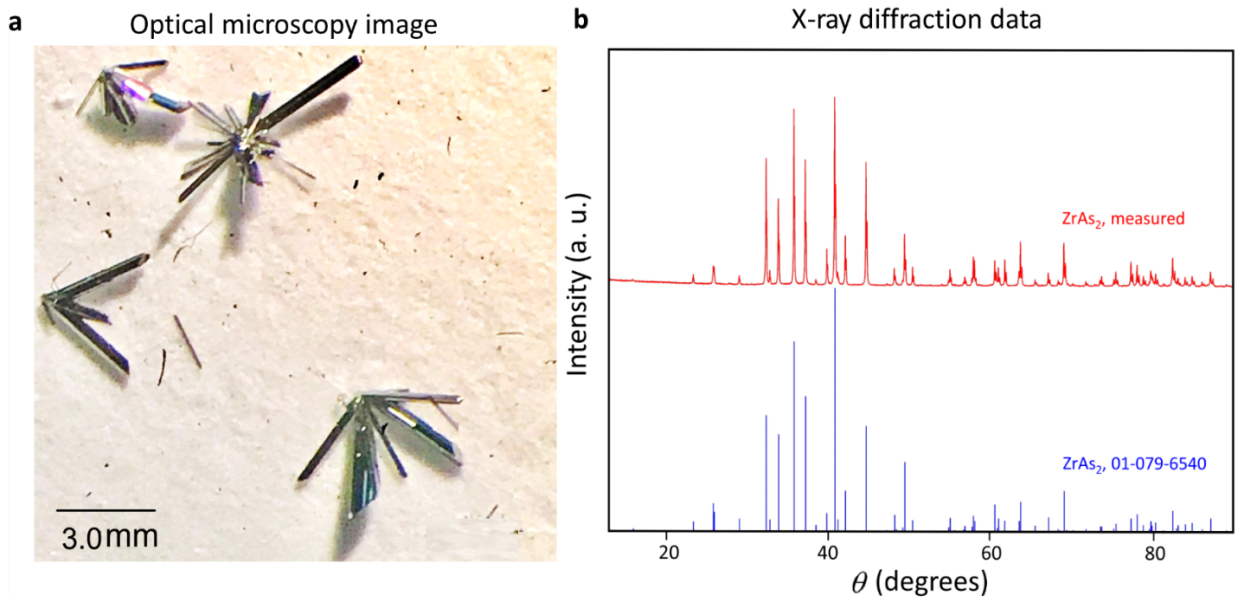

**Extended Fig. 1. Sample Characterization. a**, Optical microscopy image of the as-grown ZrAs$_2$ single crystals. **b**, Powder X-ray diffraction (XRD) profile matching the DB Card No: 01-079-6540 (ICDD (PDF-4+ 2020 RDB)). XRD data confirms the Cotunnite mineral structure for ZrAs$_2$, which crystallizes in the orthorhombic *Pnma* (62) space group.

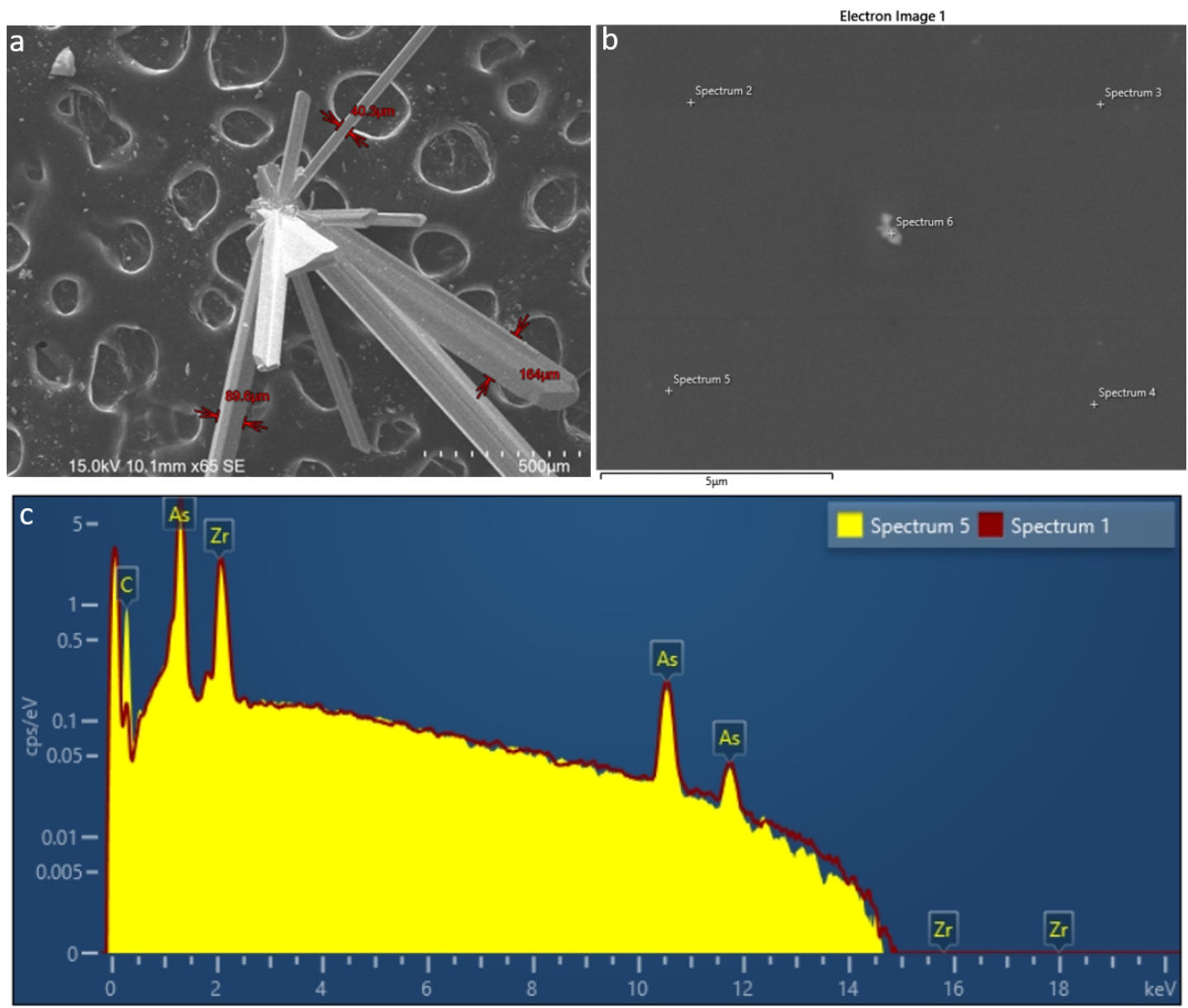

**Extended Fig. 2: EDS measurements on ZrAs$_2$ crystals. a**, SEM image of the crystals. **b**, Typical SEM image showing the arrangement of probed points within the selected area. **c**, Corresponding EDS spectrum. Various portions of the crystal shown in panel a were examined to estimate the chemical composition. The selected area reveals that, except for carbon (used to adhere the sample to the holder), only Zr and As peaks are present in the EDS spectrum.

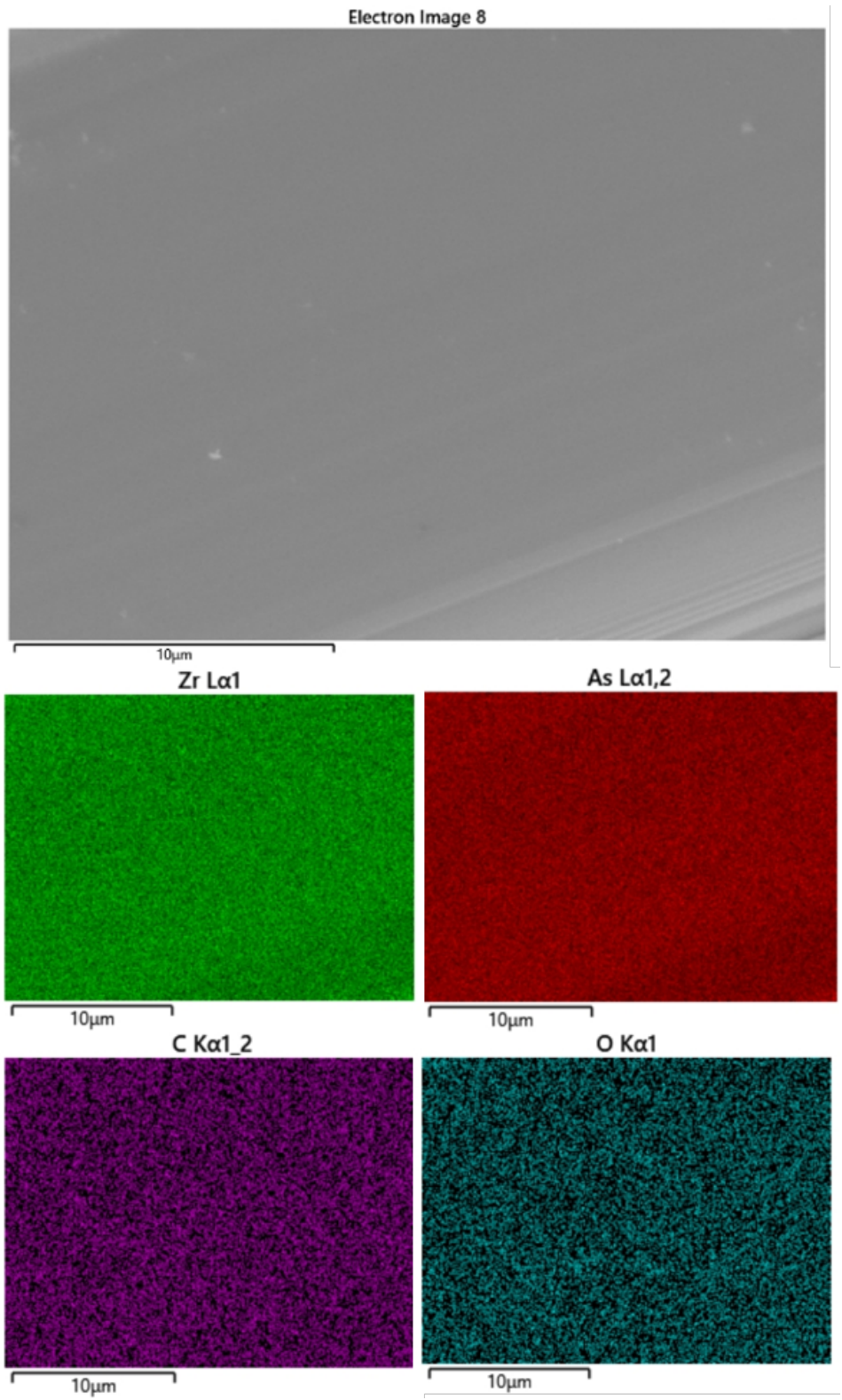

**Extended Fig. 3: EDS Mapping of the flat part of the crystal.** The mapping results clearly show only the presence of Zr and As, alongside carbon (from the substrate) and oxygen (from air).

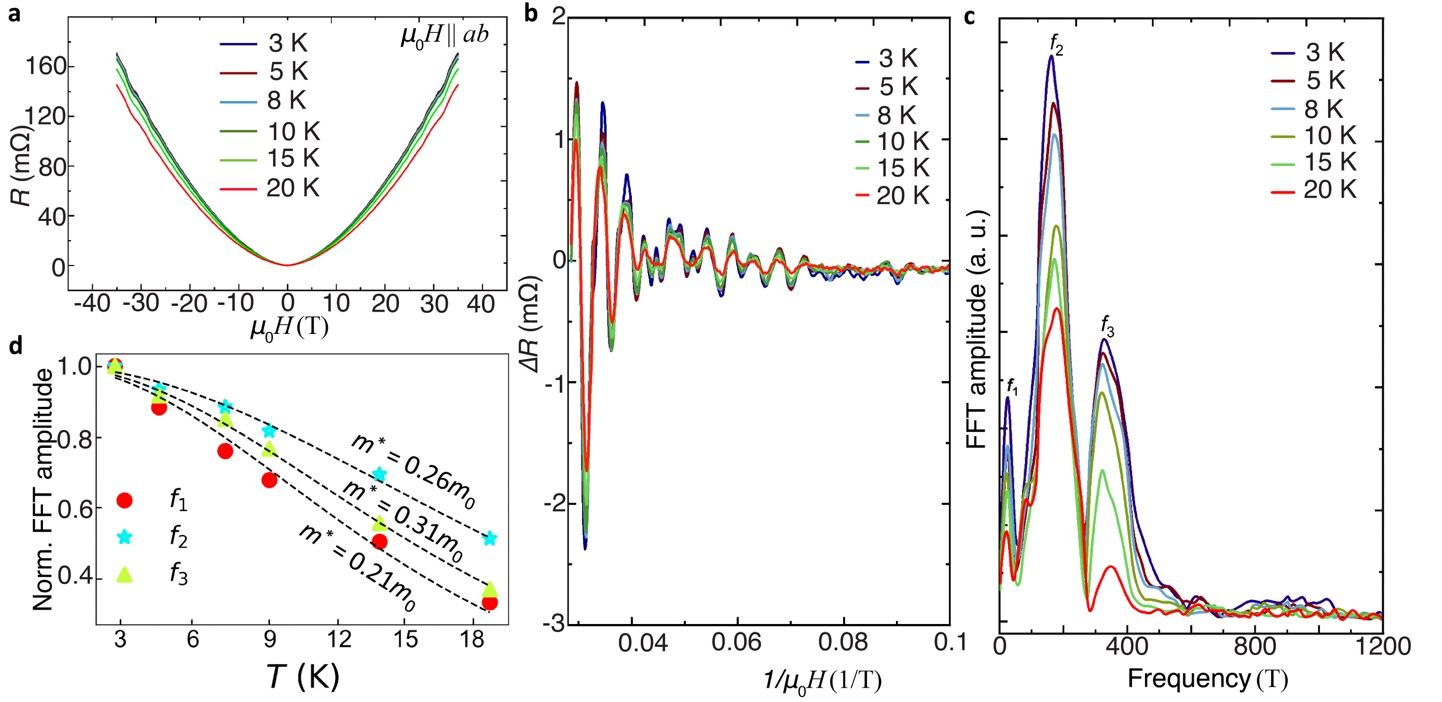

**Extended Fig. 4: Extraction of carrier effective masses from temperature dependent Shubnikov de Haas oscillations. a,** Resistance of ZrAs$_2$ as a function of the magnetic field for various temperatures. **b,** Oscillatory component ($\Delta R$) superimposed onto the magnetoresistance traces in panel **a**, revealing the gradual reduction of the quantum oscillatory amplitude as the temperature is raised. **c,** Temperature dependence of the Fourier transform of the oscillatory signal, revealing the gradual weakening of the peaks as the temperature increases. **d,** Magnitude of the Fourier transform for all three peaks (f$_1$, f$_2$, and f$_3$) as a function of the temperature. The dashed curves represent fits to the Lifshitz-Kosevich formalism, yielding carrier effective masses of 0.21 $m_e$ (f$_1$), 0.26 $m_e$ (f$_2$), and 0.31 $m_e$ (f$_3$) for three distinct Fermi surface pockets, where $m_e$ is the bare electron mass.

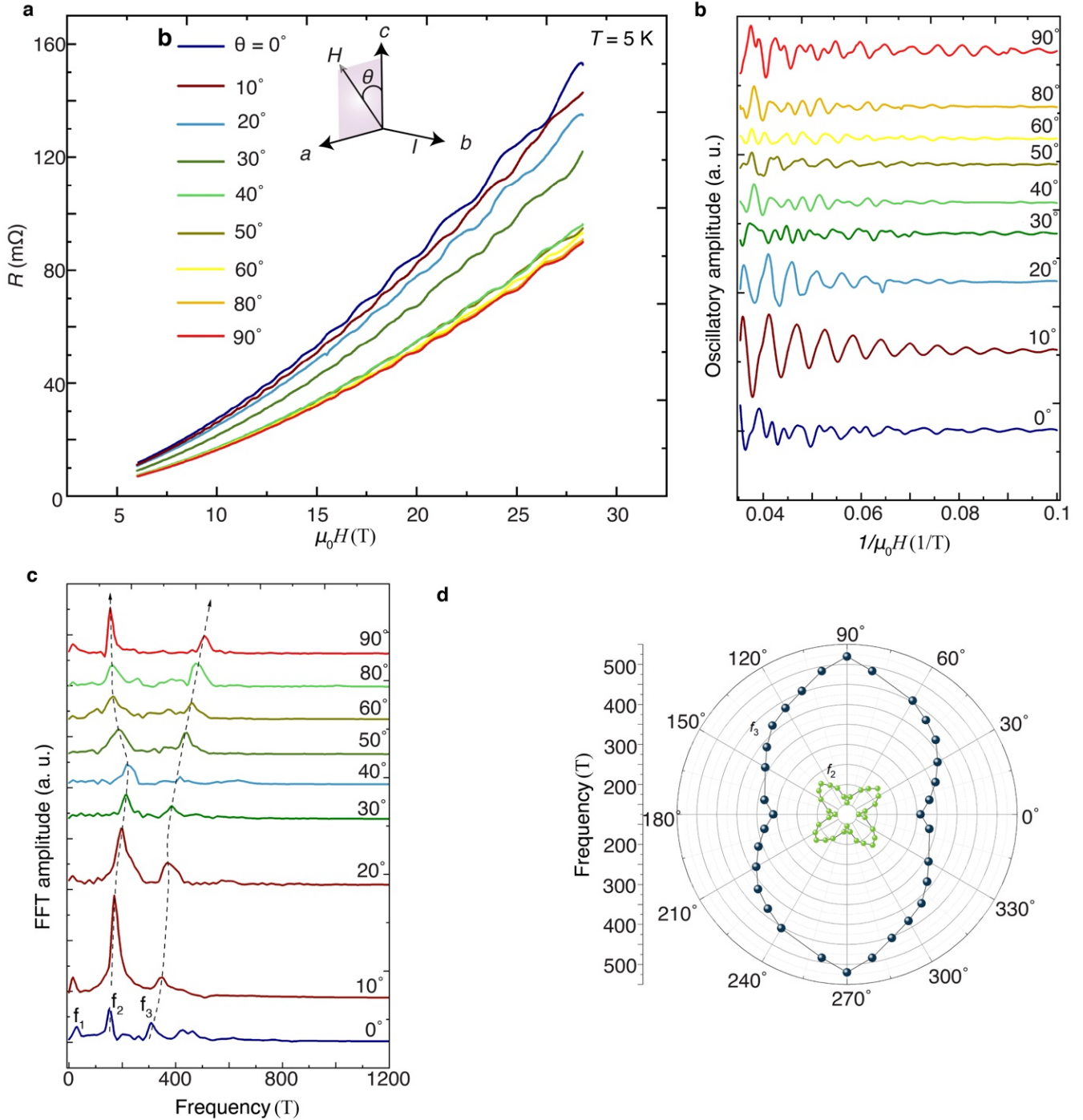

**Extended Fig. 5: Fermiology of ZrAs₂ revealed through the Shubnikov- de Haas effect. a,** Resistance as a function of the magnetic field, showcasing the superimposed oscillatory signal as the magnetic field orientation varies from the *c*- [001] ($\theta = 0^o$) to the *a*-axis [100] ($\theta = 90^o$), revealing the presence of an oscillatory signal. The inset illustrates the direction of rotation. **b**, Oscillatory component ($\Delta R$) superimposed onto the magnetoresistance traces in panel **a**. These traces were obtained by subtracting a smooth background, and were collected at various angles ($\theta$) between the magnetic field and the crystallographic *c*-axis. $\Delta R$ is plotted as a function of $1/\mu_0 H$ to reveal the $1/\mu_0 H$-periodic quantum oscillations. Traces are vertically offset for clarity, and the corresponding $\theta$ values are provided for each trace. **c**, Fourier transform spectra of the ΔR traces shown in panel **b,** with traces vertically displaced for enhanced clarity. Three distinct peaks emerge in the spectrum collected for each angle $\theta$, denoted as $f_1$, $f_2$, and $f_3$. Peak $f_2$ shifts towards higher frequencies as $\theta$ increases up to $\theta = 40°$ and then decreases. In contrast, peak $f_3$ gradually shifts to higher frequencies from $\theta = 0°$ to $90°$. **d,** Polar plot depicting the angular dependence of the peaks observed in the Fourier transforms (see also Extended Table 3). The positions of the peaks $f_2$ and $f_3$ are tracked across different angles, revealing the anisotropic, but definitively not two-dimensional, nature of the Fermi surface. These angular-dependent measurements were collected in a different cooldown and distinct magnet with respect to the $T$-dependent quantum oscillation data (Extended Fig. 4).

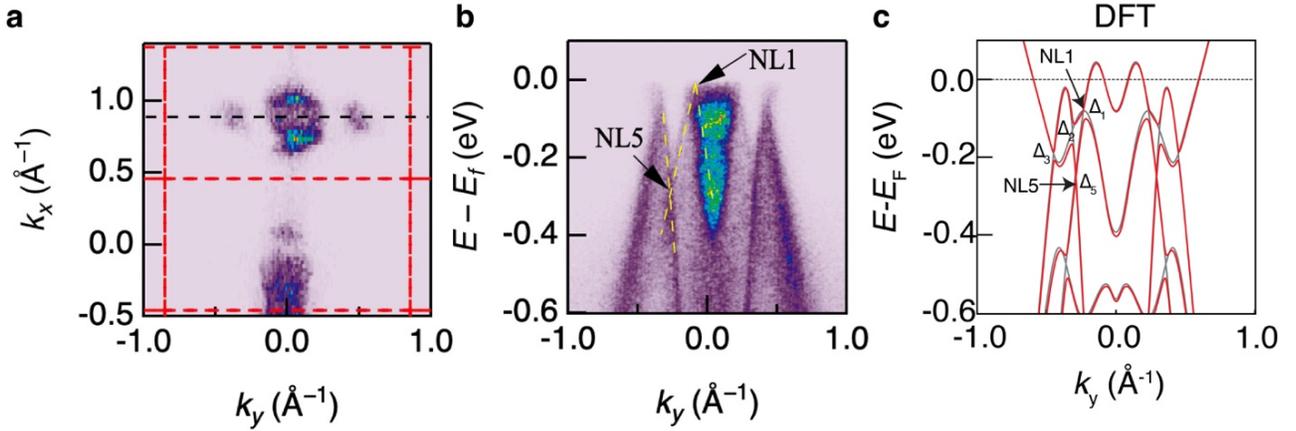

**Extended Fig. 6: ARPES evidence for nodal line crossings. a**, Fermi surface at $k_z \approx 0$, measured using 140 eV photon energy and linear horizontal polarized light. **b**, ARPES energy-momentum cut along the $\overline{Y} - \overline{\Gamma} - \overline{Y}$ path in the second Brillouin Zone, as indicated by the dashed line in a. Two sets of band crossings are observed and attributed to nodal lines 1 (NL1) and 5 (NL5), respectively, based on comparisons with the DFT calculations shown in c (see, also Fig. 1c for the naming of the nodal lines in ZrAs$_2$). Yellow dashed lines are included as visual guides for the observed band dispersions. **c**, Corresponding *ab-initio* calculations (including spin-orbit coupling) for the bulk band structure. The spin-orbit-induced gaps for the nodal lines/loops NL1, NL2, NL3, and NL5 are $\Delta_1 \approx 10$ meV, $\Delta_2 \approx 20$ meV, $\Delta_3 \approx 50$ meV, and $\Delta_5 \approx 2$ meV, respectively. $\Delta_1$ and $\Delta_5$ fall below the ARPES energy resolution (approximately 15 meV) and, therefore, cannot be observed under the presence of the $k_z$ or extrinsic broadening and thus can be generally ignored. Specifically, $\Delta_5$ is equivalent to the thermal broadening at the ARPES measurement temperatures, and thus NL5 can be considered as a nodal line state at the measured temperatures.

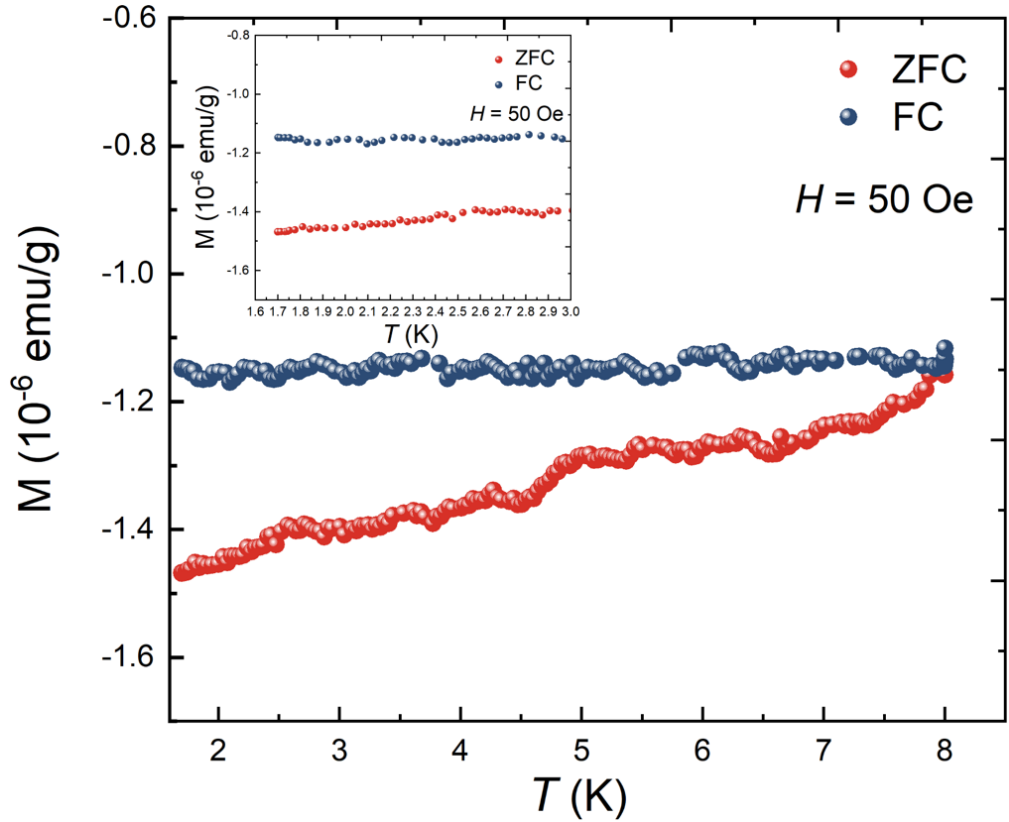

**Extended Fig. 7: Magnetic susceptibility measurements of ZrAs$_2$, highlighting the absence of a bulk Meissner effect.** Temperature-dependent magnetic susceptibility, recorded under a 50 Oe magnetic field applied along the *ab*-plane of the ZrAs$_2$ crystal, is depicted for zero-field-cooled (ZFC) and field-cooled (FC) conditions by red and blue circles, respectively. These results highlight a subtle diamagnetic response spanning the entire temperature range. The inset presents a magnified data view at low temperatures down to $T = {\sim}1.69$ K, underscoring the absence of any discernible transition or enhancement in the diamagnetic response as expected for the Meissner effect from a bulk superconductor.

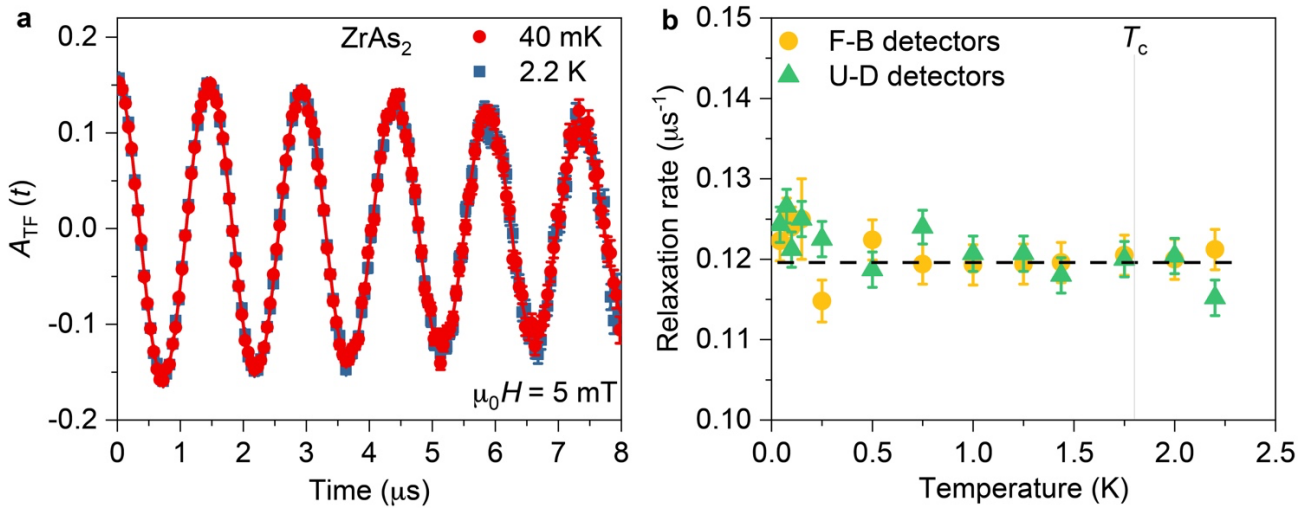

**Extended Fig. 8: Muon spin relaxation (μSR) data indicating the absence of bulk superconductivity. a**, Transverse-field (TF) μSR time spectra obtained above (2.2 K) and below (0.04 K) the superconducting transition temperature ($T_c$) for ZrAs$_2$, after field cooling the sample from above $T_c$. **b**, Temperature dependence of the muon spin depolarization rate ($\sigma$), measured down to 40 mK in an applied magnetic field of $\mu_0 H = 5$ mT for ZrAs$_2$. The temperature-independent relaxation rate across $T_c$ suggests the absence of a bulk superconducting transition.

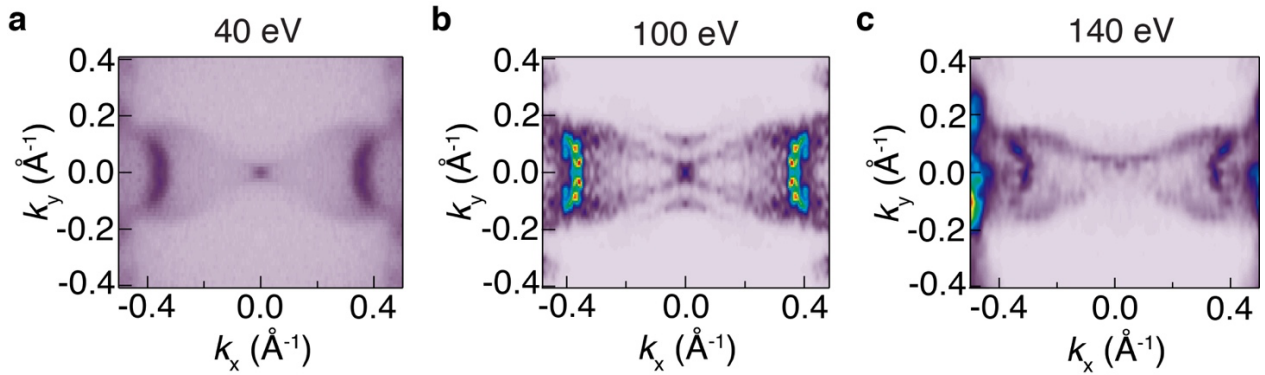

**Extended Fig. 9: Photon energy dependence of the van Hove singularity (vHs) state. a-c**, Constant energy contours at the energy of the vHs ($E_{vHs}$), measured at photon energies $h\nu = 40$, 100, and 140 eV, respectively. The shape of the butterfly-shaped pocket remains unchanged across different photon energies, suggesting that the vHs is formed by the surface state.

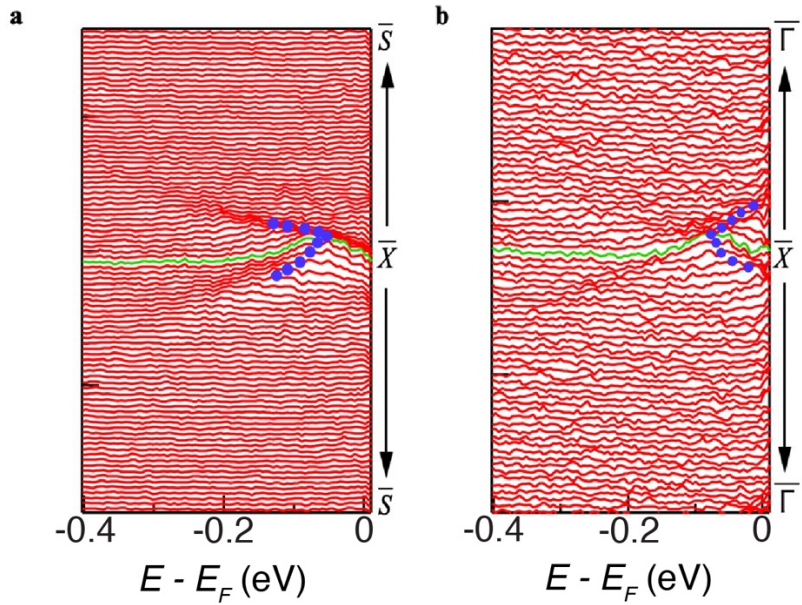

**Extended Fig. 10. Energy distribution curves (EDCs) near the surface van Hove singularity (vHs) along two orthogonal directions. a,** EDCs along the $\bar{S} - \bar{X} - \bar{S}$ path. **b,** EDCs along the $\bar{\Gamma} - \bar{X} - \bar{\Gamma}$ path. Blue circles indicate the peak positions in the EDCs.

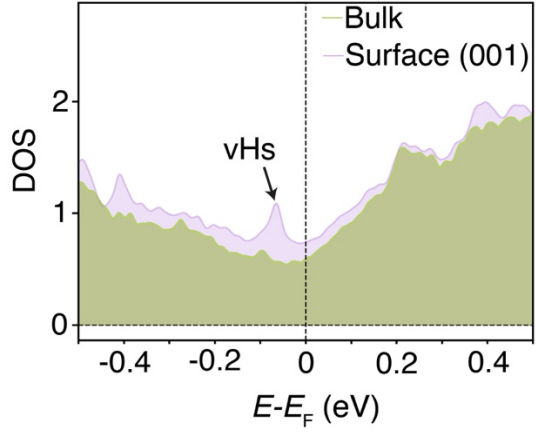

**Extended Fig. 11: Density of states comparison between (001) surface of ZrAs$_2$ and the bulk ZrAs$_2$.** *ab-initio* density of states calculated for both the (001) surface and bulk ZrAs$_2$. Notably, the ZrAs$_2$ (001) surface exhibits an elevated density of states at and near the Fermi level when compared to bulk ZrAs$_2$.

**Extended Table 1:** Mean values of As and Zr chemical content (with mean absolute deviations) for different areas (as shown in Extended Fig. 2) and five random points.

| | S1 | | S2 | | S3 | | S4 | | S5 | | S6 | |
|---|---|---|---|---|---|---|---|---|---|---|---|---|
| Chemical content | Mean | Mean absolute deviation | Mean | Mean absolute deviation | Mean | Mean absolute deviation | Mean | Mean absolute deviation | Mean | Mean absolute deviation | Mean | Mean absolute deviation |
| As (area) | 0.670 | 0.007 | 0.674 | 0.007 | 0.682 | 0.009 | 0.673 | 0.007 | 0.668 | 0.008 | 0.694 | 0.008 |
| Zr (area) | 0.330 | 0.005 | 0.326 | 0.005 | 0.318 | 0.006 | 0.327 | 0.005 | 0.332 | 0.005 | 0.306 | 0.006 |
| As (pts) | 0.686 | 0.007 | 0.682 | 0.001 | 0.703 | 0.004 | 0.688 | 0.002 | 0.690 | 0.007 | 0.703 | 0.002 |
| Zr (pts) | 0.314 | 0.007 | 0.318 | 0.001 | 0.297 | 0.004 | 0.312 | 0.002 | 0.310 | 0.007 | 0.297 | 0.002 |

**Extended Table 2:** Mean values of As and Zr chemical content (with mean absolute deviations) calculated based on all data (from Extended Table 1).

|  | S1 | |
| --- | --- | --- |
|  | Mean | Mean absolute deviation |
| As (area) | 0.677 | 0.007 |
| Zr (area) | 0.323 | 0.007 |
| As (pts) | 0.692 | 0.007 |
| Zr (pts) | 0.308 | 0.007 |

**Extended Table 3:** Peak positions of the quantum oscillatory frequencies as a function of the angle between the magnetic field and the *c*-axis (data points for Extended Fig. 5).

| Angle ($\theta$) | $f_1$ (T) | $f_2$ (T) | $f_3$ (T) |
| --- | --- | --- | --- |
| $0^0$ | 26.16 | 159.44 | 325.33 |
| $10^0$ | 18.65 | 172.88 | 343.77 |
| $20^0$ | 25.44 | 195.77 | 370.66 |
| $30^0$ | 45.65 | 206.22 | 386.11 |
| $40^0$ | 51.88 | 226.54 | 416.23 |
| $50^0$ | 71.77 | 214.33 | 447.88 |
| $60^0$ | 92.34 | 195.76 | 465.55 |
| $80^0$ | 15.54 | 169.54 | 495.56 |
| $90^0$ | 17.51 | 156.45 | 509.66 |

## Data Availability

All data needed to evaluate the conclusions in the paper are present in the paper. Additional data are available from the corresponding authors upon reasonable request.

**Competing Interests Statement**

The authors declare no competing interests.

**Acknowledgments**

We thank Sougata Mardanya, Carmine Autieri, Ruixing Zhang, and Titus Neupert for illuminating discussions. M.Z.H. group acknowledges primary support from the US Department of Energy, Office of Science, National Quantum Information Science Research Centers, Quantum Science Center (at ORNL) and Princeton University; STM Instrumentation support from the Gordon and Betty Moore Foundation (GBMF9461) and the theory work; and support from the US DOE under the Basic Energy Sciences programme (grant number DOE/BES DE-FG-02-05ER46200) for the theory and sample characterization work including ARPES. R.I and F.X acknowledges the support by the National Science Foundation under Grant No. OIA-2229498. R.I. and F.X. acknowledge the access to the computing facility Cheaha at University of Alabama at Birmingham. L.B. is supported by DOE-BES through award DE-SC0002613. Z.M acknowledge National Natural Science Foundation of China (grant no. 62150410438) and Anhui Province Key R&D Program International Cooperation Project (Grant No. 202104b11020012 The National High Magnetic Field Laboratory (NHMFL) acknowledges support from the US-NSF Cooperative agreement Grant number DMR-DMR-2128556 and the state of Florida. We thank T. Murphy, G. Jones, L. Jiao, and R. Nowell at NHMFL for technical support. M. K. would like to thank Sharif University of Technology and INSF-Grant No. 4027770.